\documentclass[twoside,leqno,twocolumn]{article}
\usepackage{ltexpprt}

\usepackage{color}
\usepackage[]{caption}
\usepackage[font=footnotesize]{subfig}
\usepackage{amsmath}
\usepackage{algorithm}
\usepackage{algorithmic}

\usepackage{floatflt}
\usepackage{graphicx}
\usepackage{url}
\usepackage{ifpdf}
\ifpdf
\usepackage{epstopdf}
\DeclareGraphicsExtensions{.eps,.pdf}
\else
\usepackage{epsfig}
\fi
\usepackage{paralist}

\usepackage{multirow}

\usepackage{xcolor}

\makeatletter
\newcommand\footnoteref[1]{\protected@xdef\@thefnmark{\ref{#1}}\@footnotemark}
\makeatother

\begin{document}

\title{\Large Accelerating Community Detection by Using K-core Subgraphs}
\author{Chengbin Peng\thanks{King Abdullah University of Science and Technology, KSA, chengbin.peng@kaust.edu.sa} \\
\and
Tamara G. Kolda\thanks{Sandia National Laboratories, CA, tgkolda@sandia.gov
}
\and
Ali Pinar\thanks{Sandia National Laboratories, CA, apinar@sandia.gov
}}
\date{}

\maketitle


\begin{abstract} \small\baselineskip=9pt
Community detection is expensive,
and the cost generally depends at least linearly on the number of vertices in the graph. We propose working with a reduced graph that has many fewer nodes but nonetheless captures key community structure.
The $K$-core of a graph is the largest subgraph within which each node has at least $K$ connections. We propose a framework that accelerates community detection by applying an expensive algorithm (modularity optimization, the Louvain method, spectral clustering, etc.)  to the $K$-core and then using an inexpensive heuristic (such as local modularity maximization) to infer community labels for the remaining nodes.  Our experiments demonstrate that the proposed framework can reduce the running time by more than 80\% while preserving the quality of the solutions. Recent theoretical investigations provide support for using the $K$-core as a reduced representation.

\end{abstract}

\section{Introduction}



Understanding community structure in a graph is a major algorithmic
challenge, with a wide spectrum of applications  across many disciplines.  Different definitions of community can be proposed,
but an essential notion is that nodes within the same community are
highly connected to one another.
Many real world networks have inherent community structure, including social networks, 
transportation networks, biological
networks, etc. \cite{girvan2002community,guimera2005worldwide,
  traud2012social}. 
Community structure has many applications, such as in network dynamics
where it is used to understand the spreading of a disease
\cite{salathe2010dynamics}. {Many aspects on community detection problems have also been explored recently, such as the impact of the degree assortativity \cite{ciglan2013community}, the contribution of node attributes 
\cite{yang2013community}, and the detection for community outliers \cite{gao2010community}.}

{There are many algorithms developed for the community detection
problem~\cite{fortunato2010community}}, and here we focus on
non-overlapping community detection without node attributes. Example methods include
Bayesian methods that maximize the likelihood of a stochastic blockmodel \cite{hofman2008bayesian}, spectral algorithms \cite{newman2013spectral}, and fast algorithms using modularity optimization \cite{blondel2008fast, le2013fast} .
These methods all share a common feature: their expense is \emph{at least} linear in the number of nodes and oftentimes much more expensive. Our goal is to work with a reduced representation that has fewer nodes but nonetheless maintains the community structure of the original.



\begin{figure}
  \centering
  \subfloat[Reduction to 4-core removes spurious edges and nodes\label{figPic}]%
  {
    \begin{minipage}[b]{1.5in}
      \includegraphics[width=1.4in]{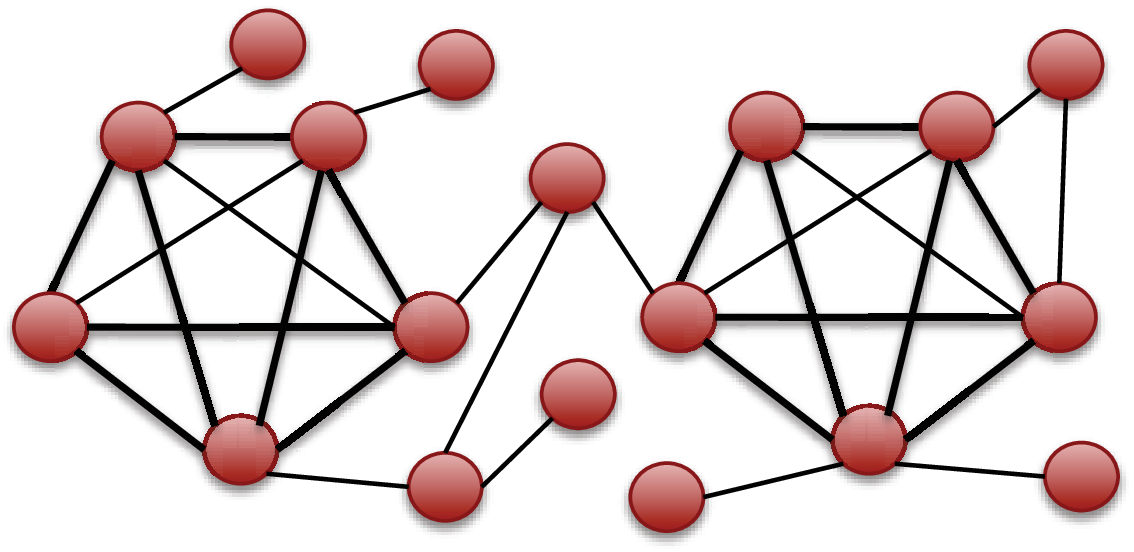}\\[2mm]
    \includegraphics[width=1.4in]{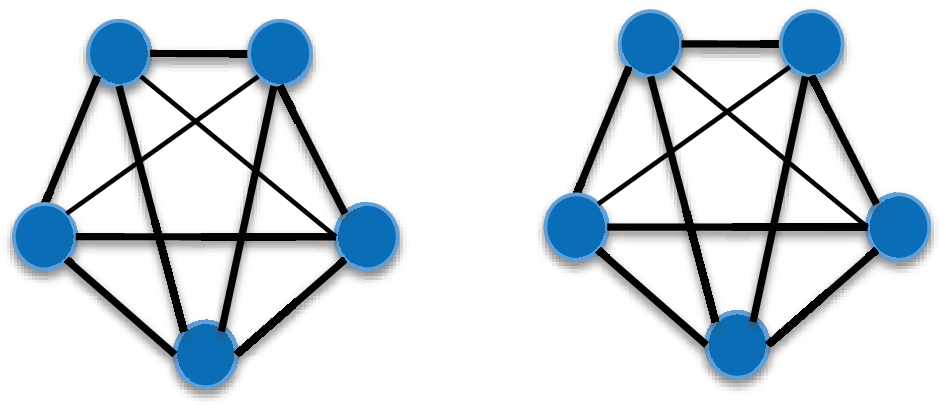}      \\[-1mm]
    \end{minipage}
}
  \subfloat[Sample results\label{figPage1}]%
{\includegraphics[width=0.25\textwidth]{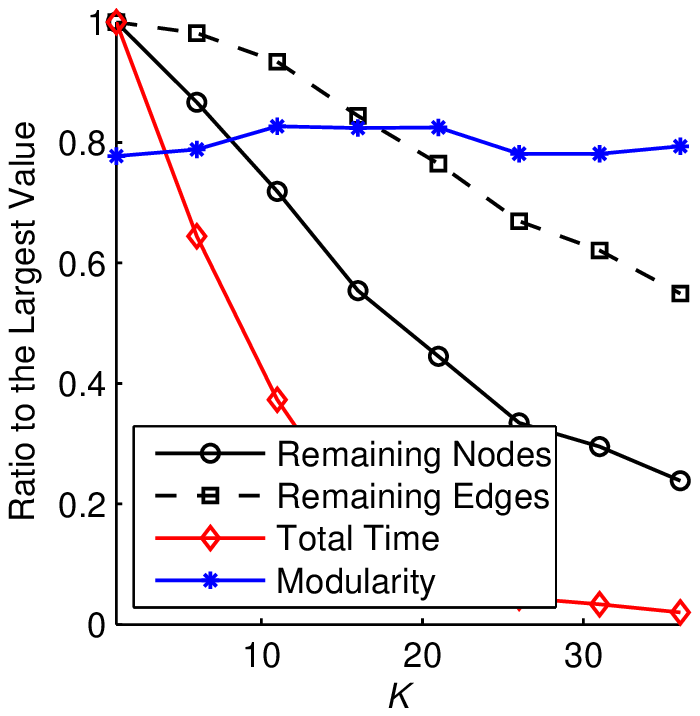}}
  \caption{Effectiveness of $K$-core subgraph community detection}

\end{figure}

The fundamental observation of this paper is that community structure is generally preserved in the $K$-core for reasonable values of $K$. Recall that
the $K$-core is the largest subgraph in which each node has at least $K$ edges \cite{dorogovtsev2006k}. For example, to compute the 4-core: remove all nodes of degree less than 4, update the degrees of all remaining nodes, and repeat until only nodes of degree 4 or higher remain. See Fig.~\ref{figPic} for an example of a graph and its 4-core.
Algorithms  for  finding  the K-core  are efficient  and amenable for parallelization \cite{sariyuce2013streaming}.
We propose to run community detection \emph{only on the much smaller $K$-core} and then use a fast heuristic to find community labels for the remaining nodes.

In Fig.~\ref{figPage1}, we show sample results for the ego-Facebook
network (4k nodes, 88k edges). This is an interesting example because the network is fairly dense, so community structure is retained for fairly large values of $K$.
We use modularity optimization \cite{newman2004fast} as an
exemplar community detection method.
For $K=1$, we see the results of applying the community detection method to the full network, and this serves as the baseline. The runtime is 876 seconds.
As we increase $K$, the number of nodes and edges decrease so that we have 20\% of nodes and 50\% of edges remaining for $K=40$. The number of edges does not decrease as quickly as the number of nodes since we are removing low-degree nodes.
Because modularity optimization is an expensive procedure, the time reduces faster even than the number of nodes, to less than 20\% of the total time for the baseline.
The number of communities is determined automatically, but the highest number of communities found in this example is 14.
%
We measure the quality in terms of modularity. 
In this case, the solution is always at least as good as the baseline.

\subsection{Our Contributions}


\begin{compactitem}
  \item \textbf{General $K$-core Community Detection Framework}: We propose a fairly intuitive but nonetheless novel method for accelerating community detection by focusing the bulk of the effort on the $K$-core. We suggest a fast heuristic for providing labels to nodes outside the $K$-core.

\item \textbf{Time Reduction \& Quality Preservation}: We show positive results (similar to Fig.~\ref{figPage1}) for a variety of popular community detection algorithms on various real-world data sets. Overall, we find that we get substantial decreases (up to 90\%) in runtime without sacrificing quality, measured in terms of modularity, normalized cut, and NMI.

\item \textbf{Theoretical Justification}:  We show how recent theoretical developments support the idea that the $K$-core preserves the community structure and also how this understanding can be used to select a good value for $K$.
\end{compactitem}

\subsection{Related Work}\label{sectionRelatedWork}

Much research has been devoted to community detection algorithms \cite{leskovec2010empirical, fortunato2010community}. However, scalability is a major issue. One strategy  is to split the graph into very small subgraphs and optimize each small problem locally \cite{raghavan2007near, huang2011towards}, in which case the quality is usually compromised.  Alternatively, another strategy is to include additional phases for global optimization  \cite{le2013fast, blondel2008fast, gargi2011large}, but the running time becomes longer.

Perhaps the most closely related work to our focuses on
removing unimportant edges \cite{satuluri2011local}. However, this method preserves all the nodes and at least one edge per node, so it requires input graphs with relatively high average degree to achieve a good performance.

Scalability can also be achieved via high-performance implementations such as multi-threaded algorithms \cite{riedy2012scalable, bu2013fast, zhu2008scalable}. Some other fast algorithms rely on additional information, such as node attributes \cite{ruan2013efficient, steinhaeuser2008community}. Our $K$-core framework is compatible with these improvements.

 Some papers discuss the existence condition of a $K$-core subgraph \cite{janson2007simple, riordan2008k}.  Some study the relationship between the $K$-core and the percolation dynamics \cite{serrano2006clustering2}, and it is worth noting that the bond percolation is quite similar to the label propagation in community detection  \cite{raghavan2007near}. Recently, $K$-core is also used to evaluate the robustness of communities, which reflects the collaborative nature in co-authorship graphs \cite{giatsidis2011evaluating}.



\section{Algorithm} \label{sectionAlgo}


%
Our framework is compatible with any community detection method and  comprises three steps. The first step is to reduce the whole graph to a $K$-core. The second step uses an existing algorithm to generate community labels for nodes in  the $K$-core. The third step is to use find community labels for the remainder of the graph via a fast algorithm.
Algorithm \ref{alg1} is the pseudocode of the framework; here, $G$ is the original graph, $K$ is the desired core number, $G_K$ is the $K$-core subgraph, $g_K$ is the labels for the $K$-core, and  $g$ is the labels for all nodes.
Note that we discuss the choice of $K$ in Sec.~\ref{sectionTheory}.

\begin{algorithm}
\caption{Accelerated $K$-core Community Detection}
\label{alg1}
\begin{algorithmic}   [1]
    \STATE  \textbf{input} Graph $G$, Parameter $K$
    \STATE  \textbf{output} Labels $g$
    \STATE (1) $G_{K} \leftarrow$  Kcore($G$, $K$)
    \STATE (2) $g_{K}\leftarrow$  CommunityDetection($G_{K}$)
    \STATE (3) $g\leftarrow$ Recover($G$, $g_{K}$)
    \RETURN $g$
\end{algorithmic}
\end{algorithm}

\subsection {Find the K-core Subgraph}
There are many algorithms available to find the $K$-core of a graph \cite{sariyuce2013streaming, janson2007simple}. A standard approach \cite{janson2007simple}
is to traverse all the nodes and remove those connected by less than $K$ edges. The edges connected to the removed nodes are removed as well. The traversal and removal may be repeated multiple times, until all the remaining nodes have degrees at least $K$.
The pseudocode is presented in Algorithm \ref{algKcore1}.

\begin{algorithm}
\caption{Algorithm to Find the $K$-core Subgraph}
\label{algKcore1}
\begin{algorithmic}   [1]
    \STATE  \textbf{input} Graph $G$, Parameter $K$
    \STATE  \textbf{output} Subgraph $G_{K}$
   \STATE $G_{K} \gets G$
  \WHILE {$G_{K}$ is not a $K$-core}
	\STATE Find nodes in $G_{K}$ whose degree is less than $K$
	\STATE Remove those nodes and their incident edges
	\STATE Update the node degrees for the remaining nodes
  \ENDWHILE

    \RETURN $G_{K}$
\end{algorithmic}
\end{algorithm}

\subsection {Apply a  Community Detection Algorithm  on the K-core}

Applying a community detection algorithm to the $K$-core is the main part in our framework,
and any existing algorithm can be used here. The biggest performance gains will be for algorithms with high quality and high time complexity.

The main algorithm detects the community structure only for nodes in the $K$-core, which is the main reason for the running time reduction.
Here we discuss briefly the effect of our framework on different algorithms. For algorithms with high quality and fairly fast running time, our framework can reduce the running time significantly with the quality preserved. Those algorithms include the Louvain method \cite{blondel2008fast}, Martelot's method \cite{le2013fast}, and modularity optimization \cite{newman2004fast}. For very fast algorithms such as label propagation \cite{raghavan2007near}, our framework does not improve the running time except for very large graphs. Some algorithms like the Bayesian method \cite{hofman2008bayesian} and spectral clustering \cite{von2007tutorial} need to specify a community number before running, but our framework can increase the community number (in the ``recover'' step) from the predefined value and to improve the quality.

We note that any method may be used for community detection, though each may rely on a different merit function. This is discussed further in the next subsection.

\subsection{Recover the Community Structure for the Whole Graph} \label{sectionRecoveryPro}

When the community structure $g_K$ of the $K$-core   is known, the third step is to recover the node labels for all the nodes in and outside the $K$-core. With the knowledge of  $g_K$, it becomes possible to use a simple algorithm to achieve high quality.
In fact, the exact details of the recovery algorithm do not have much impact on the overall success of the method. Low-degree nodes are generally not critical for community structure.

The recovery step is composed of two stages: inferring community labels for nodes outside the $K$-core and optimizing the structure for the whole graph, as shown in Algorithm \ref{algRecover1}.

\begin{algorithm}
\caption{Algorithm to Find the Community Structure of the Whole Graph}
\label{algRecover1}
\begin{algorithmic}   [1]
    \STATE  \textbf{input} Graph $G$, Labels $g_{K}$
    \STATE  \textbf{output} Labels $g$
    \STATE $s \gets $ set of unlabeled nodes, roughly sorted in descending order according to the proportion of their neighbors inside the $K$-core
    \REPEAT
	 \FORALL {nodes $i \in s$}
         \STATE $g(i) \leftarrow ${plurality vote of labeled neighbors of $i$} 
         \STATE Remove $i$ from $s$
	\ENDFOR	
	\UNTIL{no more nodes can be labeled}
    	\STATE $g \leftarrow $ SecondStageAlgorithm$(A, g)$
    \RETURN $g$
\end{algorithmic}
\end{algorithm}

In the first stage, we sort the unlabeled nodes in a descending order according to the ratio of their labeled neighbors. Sorting is necessary to make sure that the community structure information of the $K$-core is sufficiently utilized,
so it runs at the beginning of this stage.

A sorting algorithm typically has a time complexity of $O(n\log n)$. However it can be reduced to $O(n)$ by using a coarse sorting instead. Since the ratio of neighboring nodes is in the interval of $[0,1]$, in coarse sorting we can divide the interval into a number of evenly-sized bins and sort accordingly. For instance, if we have 10 bins where bin 1 has the highest ratio (i.e., (0.9,1]) and bin 10 has the lowest ratio (i.e., [0,1.0]) the nodes with ratios 0.85, 0.5, and  0.45 and put into bins 2, 6, and 6, respectively.


By the order from the sorting above, we assign community labels of the unlabeled nodes according to their most frequent neighboring label (ties are broken randomly). Each assignment may impact subsequent labels.  Additionally, we repeat the procedure (without resorting) until we have labeled every connected node.
Any nodes that are not connected to the $K$-core are unlabeled. Each unlabeled node is put into its own single-node community.

In the second stage, we optimize the community structure for the whole graph. Although it can be fast by repeatedly adjusting the labels of nodes according to their most frequent neighboring labels using label propagation  \cite{raghavan2007near}, we propose a local modularity optimization algorithm to produce a better quality. This method updates the labels according to the modularity gain by the following analysis.

The modularity \cite{clauset2004finding} is defined as  follows. Let $A$ be the adjacency matrix of $G$ and $m$ be the number of edges. Define $e_{r} = \frac{1}{2m}\sum_{ij}A_{ij}\delta(g(i),r)\delta(g(j),r)$ to be the fraction of edges that join vertices within community $r$ and $a_{r} = \frac{1}{2m}\sum_{ij} A_{ij} \delta(g(j),r)$ to be the fraction of edges that are attached to vertices in community $r$. The modularity is
\begin{equation}
Q = \sum_r (e_r - a_{r}^2).
\end{equation}
%
Therefore, when a node $i$ is moved from community $r$ to community $s$ ($r \neq s$), the change of modularity is as follows.
\begin{align}
\Delta Q =& Q^{(2)} - Q^{(1)}\nonumber\\
= & \sum_i [e_i^{(2)} - (a^{(2)}_{i})^2] - \sum_i [e_i^{(1)} - (a^{(1)}_{i})^2] \nonumber\\
= & [(e_r^{(1)} - d_r) - (a^{(1)}_{r} - d)^2] \nonumber \\
&+ [(e_s^{(1)} + d_s) - (a^{(1)}_{s}+d)^2]\nonumber\\
  & - [e_r^{(1)} - (a^{(1)}_{r})^2] - [e_s^{(1)} - (a^{(1)}_{s})^2]\nonumber\\
=& (- d_r+d_s) + 2d(a^{(1)}_{r}-a^{(1)}_{s}) - 2d^2.
\end{align}
where $d_r = \frac{1}{2m}[\sum_{j \neq i} (A_{ij}+A_{ji})\delta(g(j),r)  + A_{ii}]$
 and $d = \frac{1}{2m}\sum_{j} A_{ij}$. For undirected graphs without self-loops, $d_r = \frac{1}{m}\sum_{j} A_{ij}\delta(g(j),r) $.

In local modularity optimization, we traverse all the nodes for multiple rounds, and compute the potential  $\Delta Q$ for each node over all the possible community labels. If the maximum potential $\Delta Q$ for a node is larger than zero, we change the community label of that node correspondingly; otherwise, the label of the node remains unchanged.

There is somewhat of a mix of objectives in the different steps of our method. In Step 2, the community detection optimizes according to the selected method. But in Step 3, the recovery first uses label propagation as a criteria and then switches to optimizing modularity.
It is certainly possible to make all the steps consistent, but we have selected very inexpensive heuristics for Step 3. Perhaps surprisingly, our experiments show that quality is not negatively impacted by this mixture of objective functions. It may be due to the fact that they are different ways of measuring the same fundamental structure.

\subsection{Time Complexity} \label{subsectionSpeedup}


The running time of the algorithm in the first and third step 
grows linearly with respect to  the size of the graph.
%
%
Let $b$ be some constant. Let the graph size be $N$, and a corresponding $K$-core subgraph size be $N_K$. If a community detection algorithm has time complexity $O(N^b)$
to find the community structure, it only cost $O(N_K^b)$ in the second step of our framework. 
Thus, the complexity  of the algorithm using our framework is reduced to $O(N_K^b+N)$.
Therefore, the speedup is $(\frac{N}{N_K})^b$ asymptotically.



\section{Experiments}\label{sectionExperiment}


In this section, we demonstrate that, on both synthetic and real world data,  our approach can reduce the computing time significantly, while the quality of the detected communities changes little and sometimes even becomes better.


\subsection{Data}



The synthetic graph (LFR\_N1e4) is generated by LFR benchmark  \cite{lancichinetti2009benchmarks} using the following parameters: 10000 nodes with average degree 10 and maximum degree 50, mixing parameter of 0.1, exponent of degree distribution = -2, exponent of community size distribution = -1, and the minimum and maximum community sizes as 20 and 200.


The real world graphs listed in Table \ref{tableData} are  from the Stanford Large Network Dataset Collection at SNAP\footnote{SNAP: Stanford Network Analysis Platform, \url{ http://snap.stanford.edu/}}. The ego-Facebook network also has community assignments available.

\begin {table}
\footnotesize
\caption {Data Sets} \label{tableData}
\begin{center}
\begin{tabular}{l*{3}{c}r}
Data Set Name             & Number of Nodes & Number of Edges \\
\hline
com-Youtube &1,134,890	 &2,987,624\\
com-Amazon  &334,863	&925,872  \\
Email-Enron           &36,692	&367,662  \\
ego-Facebook	&4,039	&88,234\\
ca-AstroPh &18,772	&396,160	\\
ca-CondMat &23,133	&186,936\\
ca-GrQc	&5,242	&28,980\\
ca-HepPh	&12,008	& 237,010	\\
ca-HepTh &9,877  &51,971\\
oregon1\_010331 &10,670 &22,002\\
oregon1\_010421 &10,859 &22,747\\
oregon1\_010428 &10,886 &22,493
\end{tabular}
\end{center}
\end{table}

\subsection{Implementation Details}

All the results are produced in Matlab R2013b on a Dell Precision T7500 with Dual Intel Xeon X5650 2.66GHz Six-core CPU and 48GB of RAM. The results are averaged from five runs.

To demonstrate the effectiveness and flexibility of our framework, we provide results for a variety of community detection algorithms. The details of the methods and implementations is given below. The methods automatically determine the number of communities except for the last two as noted.

\begin{compactitem}
  \item \textit{Louvain method} \cite{blondel2008fast}: 
Initializes each node as a single community, and shifts the community label of each node according to the modularity gain, until the labels converge. Then, it considers each community as a node to merge some of them again according to the modularity gain. We use the MATLAB implementation called \texttt{cluster\_jl} by A.~Scherrer with default settings\footnote{ \url{http://perso.uclouvain.be/vincent.blondel/research/Community_BGLL_Matlab.zip}}.

  \item \textit{Martelot's method} \cite{le2013fast}: 
A multi-scale algorithm with two phases, optimizing a global and a local criterion respectively. In our experiment, we use the modularity criterion. We use the MATLAB implementation called \texttt{fast\_mo} by E.~Le~Martelot\footnote{\url{http://www.elemartelot.org/index.php/programming/cd-code}\label{fn:mart}}.

  \item \textit{Modularity optimization} \cite{newman2004fast}: 
An agglomerative method that merges nodes into bigger and bigger communities hierarchically, using the modularity criterion.  We use the MATLAB implementation called \texttt{fast\_newman} by E.~Le~Martelot\footnoteref{fn:mart}.

   \item \textit{Label propagation} \cite{raghavan2007near}: 
This is a fast method that simply updates the label of a node according to the plurality vote of its neighbors (according to their labels, randomly breaking ties). It runs until the labels cease changing (or 1000 iterations). It is
appropriate for large networks, though the quality is usually compromised. Also, the update order impacts the results, so in the experiments run for five times using random orders, and the results are averaged.
We implemented this algorithm in MATLAB ourselves.

  \item \textit{Bayesian method} \cite{hofman2008bayesian}: 
A variational approach solves the parameter inference problem in a stochastic block model.
We use the MATLAB implementation called \texttt{vbmod\_restart} by J.~Hofman\footnote{http://vbmod.sourceforge.net/}. In this case, the number of communities  must be specified by the user, which is a disadvantage but not relevant for our discussion here. We use the results of the prior experiments to come up with reasonable starting values\footnote{%
Specified number of communities:
Email-Enron=485, ego-Facebook=10, ca-AstroPh=604, ca-CondMat=1835, ca-GrQc=391, ca-HepPh=608, ca-HepTh=904, oregon1\_XXXXXX=30.
\label{fn:guess}
}.%
We set \texttt{opt.NUM\_RESTARTS} to be one (default is ten) in order for the code to run relatively quickly.

  \item \textit{Spectral clustering} \cite{von2007tutorial}: 
Considers the graph as a similarity matrix, and solves a data clustering problem where each cluster is a community. It contains two phases. First, it maps the data onto a lower dimensional space formed by eigenvectors of the Laplacian matrix; second, it uses k-means to cluster the reduced data. We use the MATLAB implementation \texttt{SpectralClustering} by I.~Buerk\footnote{\url{http://www.mathworks.com/matlabcentral/fileexchange/34412-fast-and-efficient-spectral-clustering/content/files/SpectralClustering.m}} with the Jordan/Weiss normalization scheme (\texttt{type=3}), except we have modified it to run the K-means step ten times and choose the solution that minimizes the sum of the within-cluster sums of point-to-centroid distances.
Like for the Bayesian method, the user needs to specify an appropriate number of communities in advance, and we use the same values again\footnoteref{fn:guess}.

\end{compactitem}

In the recovery step of our algorithm (Step 3), the maximum number of iterations for labeling nodes is set to ten, and the maximum number of modularity optimization iterations is set to 10.

\subsection{Results and Analysis --- Synthetic Data}

%

\begin{figure*}
\centering
\begin{tabular}{cc}
 \subfloat[Graph size and details of relative run time for Louvain method \cite{blondel2008fast}]{\includegraphics[width=0.3\textwidth]{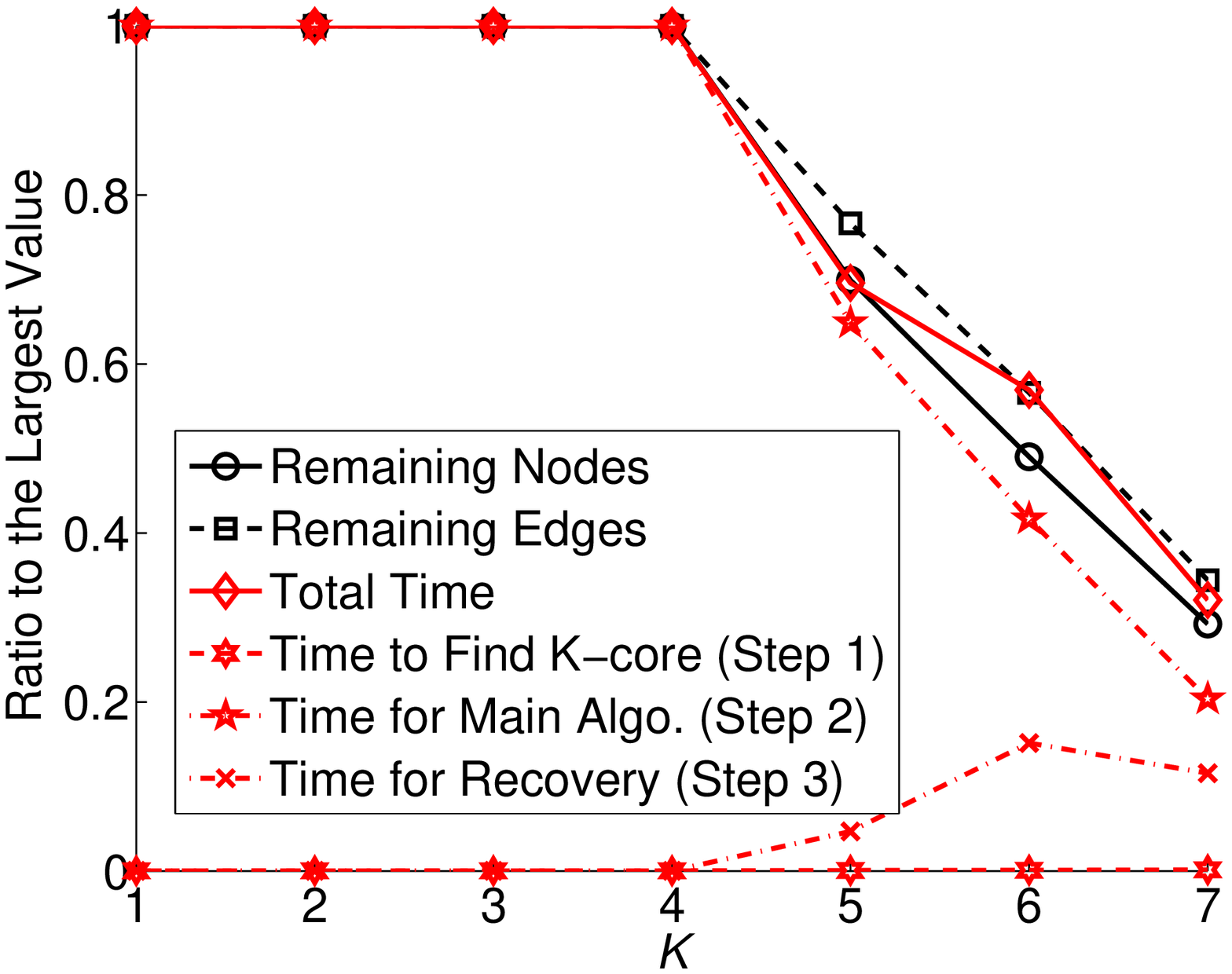}\label{figLFR1}}
&
\subfloat[Relative run time]{\includegraphics[width=0.3\textwidth]{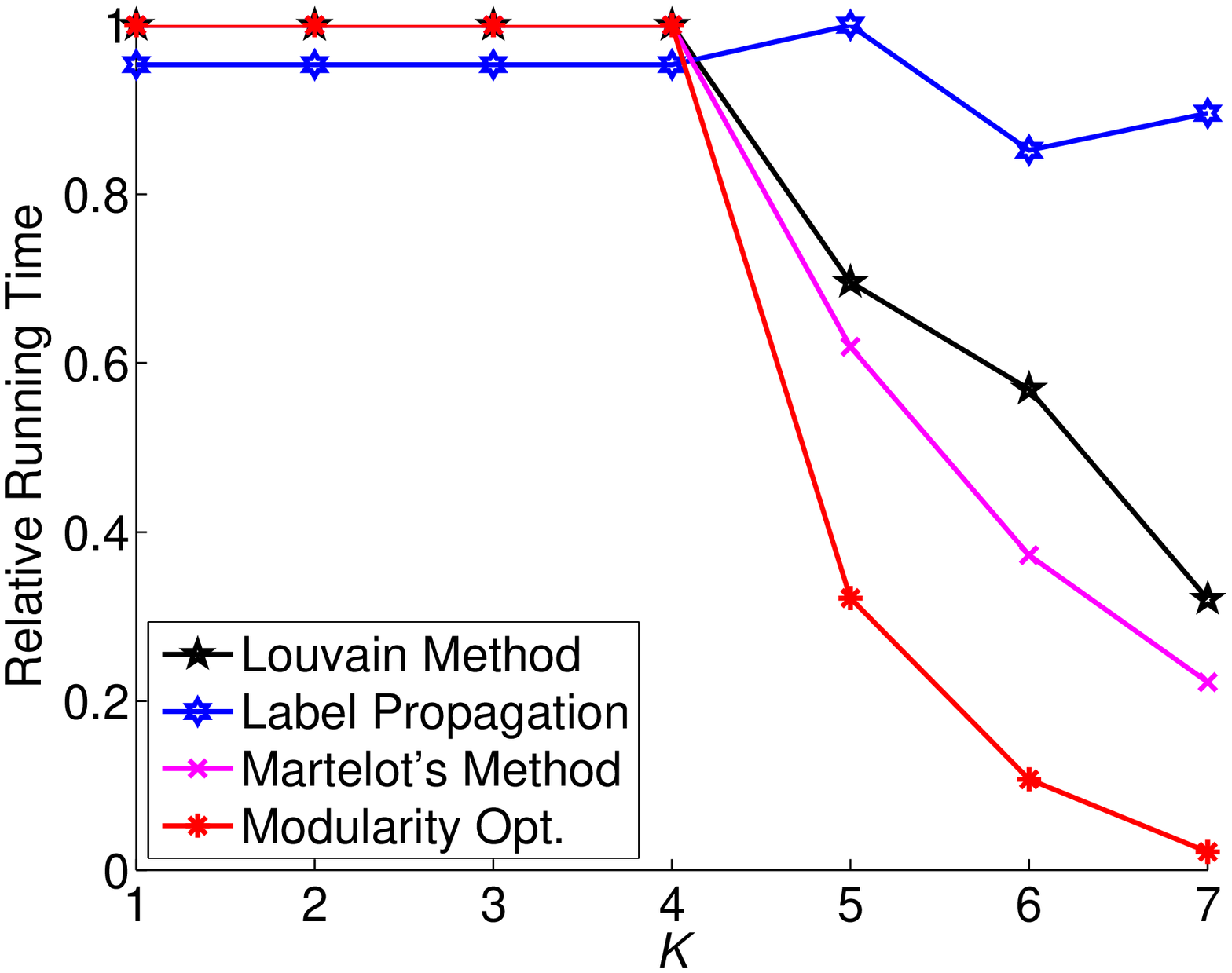}\label{figLFR3}}
\end{tabular}
\begin{tabular}{ccc}
\subfloat[Modularity]{\includegraphics[width=0.3\textwidth]{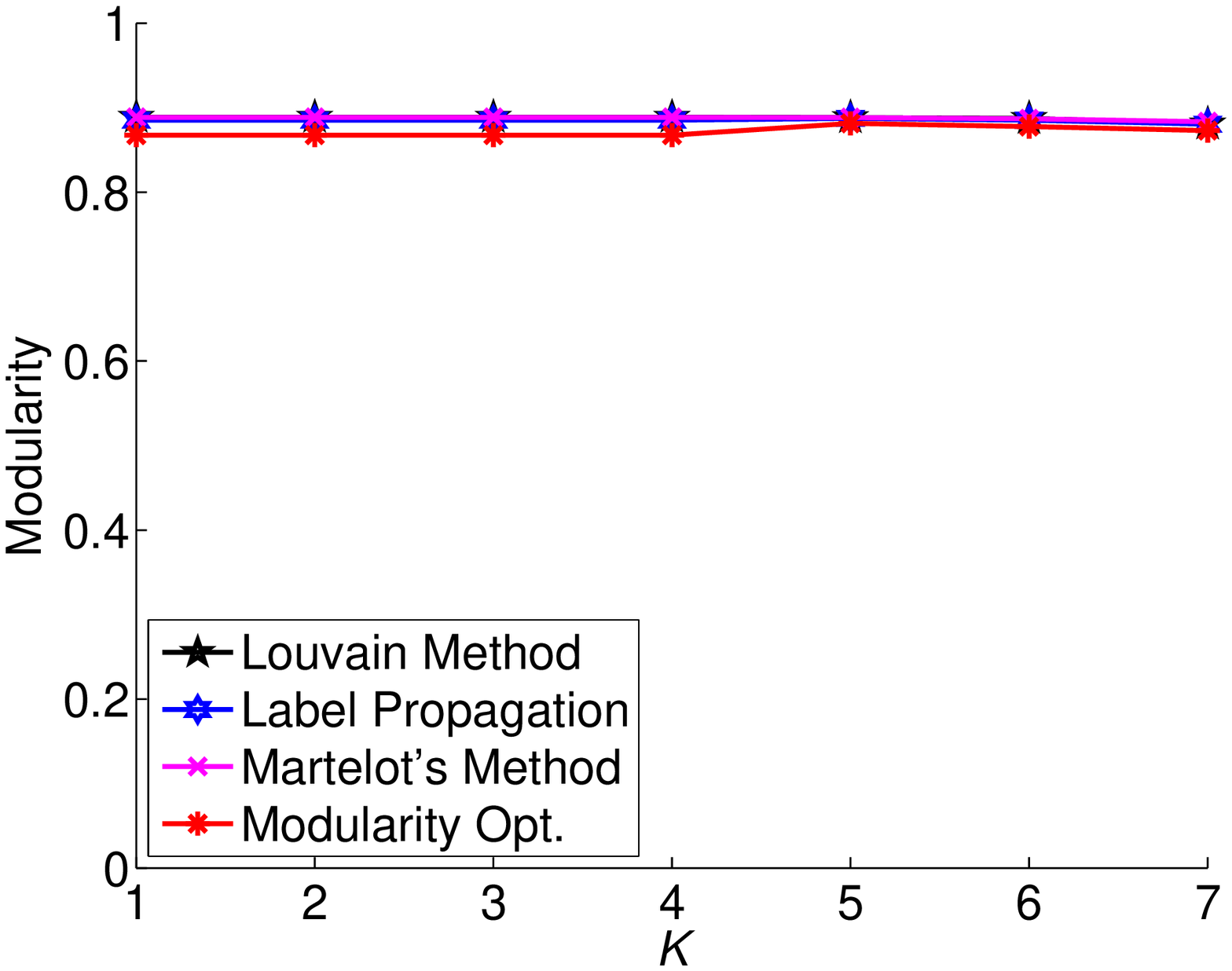}\label{figLFR2}}
&
\subfloat[NMI w.r.t.\@  ground truth]{\includegraphics[width=0.3\textwidth]{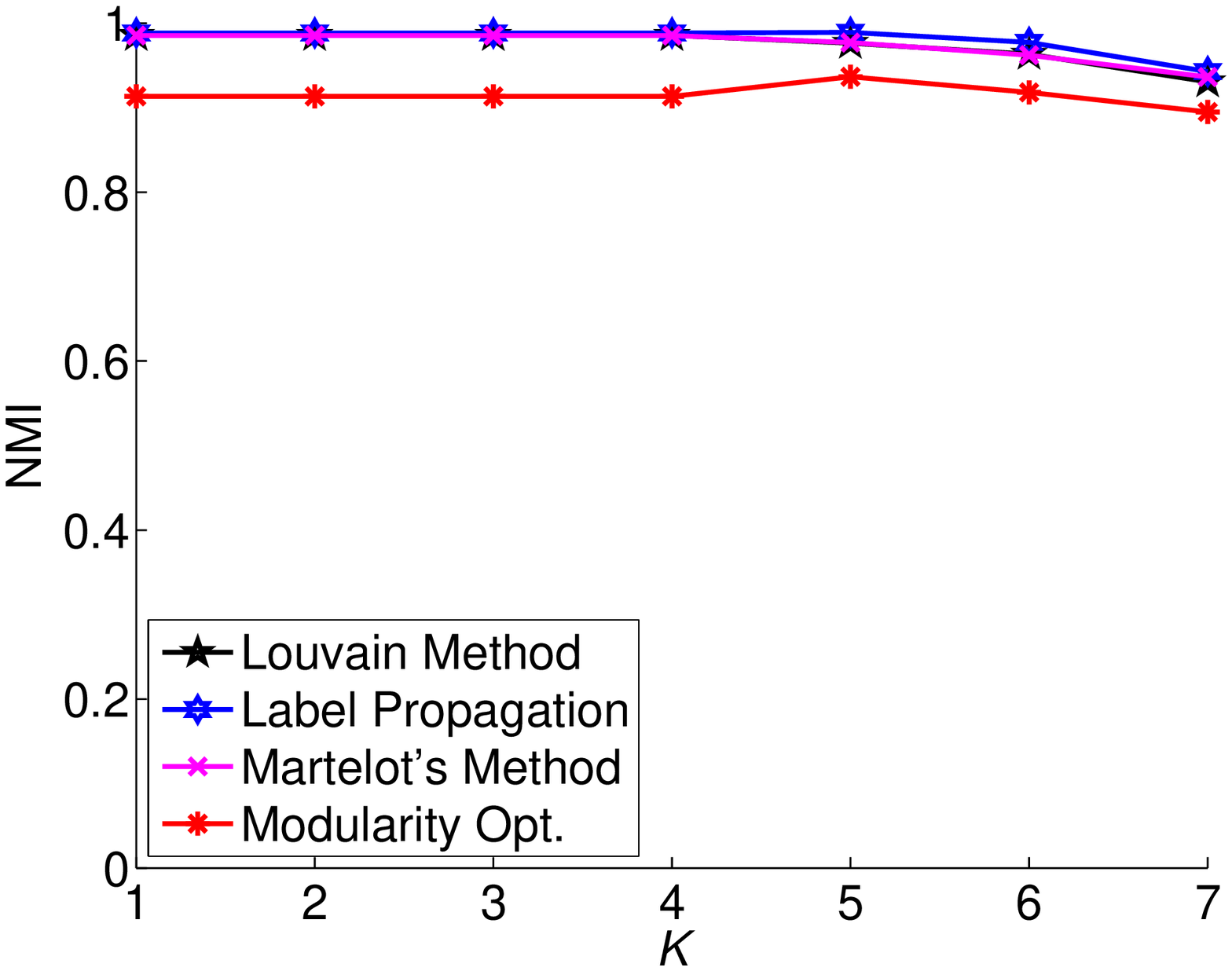}\label{figLFR4}}
&
\subfloat[NMI w.r.t.\@  baseline result]{\includegraphics[width=0.3\textwidth]{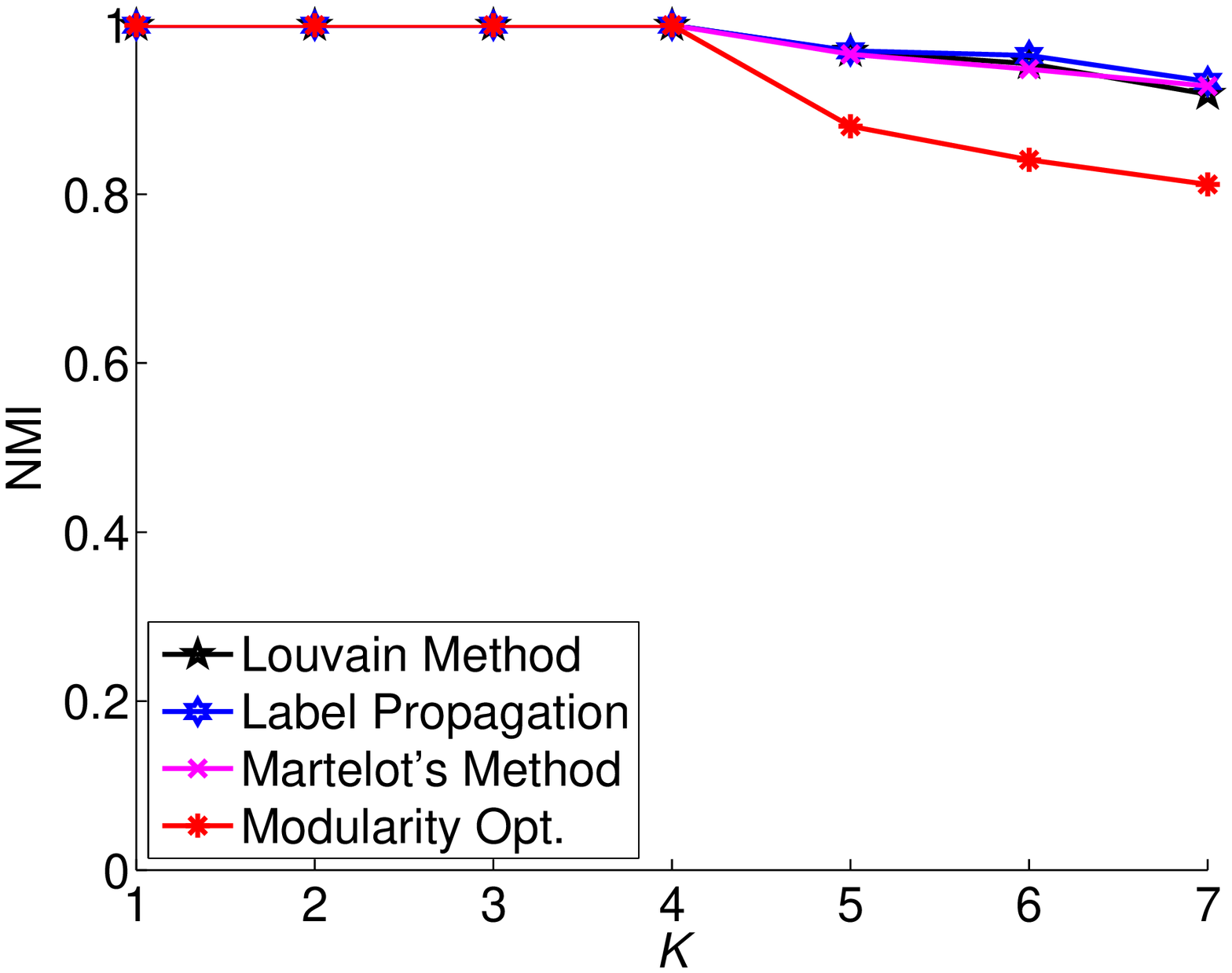}\label{figLFR5}}
\end{tabular}
\caption{Results of $K$-core acceleration on LFR\_N1e4 (artificial LFR network)}
\label{figLFR}
\end{figure*}




Fig.~\ref{figLFR} shows performance analysis using the synthetic LFR graph. In this graph, the minimum degree is 4; therefore, the 4-core is the same as the 1-core which is the original graph, so all the results for $K \leq 4$ are identical.

Fig.~\ref{figLFR1} provides a few different bits of information. The black lines show the proportion of nodes and edges in the $K$-core as $K$ increases.
The 7-core contains only 40\% of the nodes and edges of the complete graph.
The red lines in Fig.~\ref{figLFR1} give a sense of the breakdown of the runtime for a specific method (Louvain method \cite{blondel2008fast}). The diamonds show the time relative to the longest time, in this case for the full graph. The other lines show the proportion of that time that went to each step. Observe that Step 2 takes the majority of the time because it has higher complexity, whereas Steps 1 and 3 take only a small proportion of the time, growing as the number of nodes outside the $K$-core increases.


Fig.~\ref{figLFR3} shows the total run time for four different algorithms in Step 2. Here we only consider the methods that automatically determine the number of communities.
The run times are relative to the longest run time for that method and are not directly comparable \emph{across} the different methods.
When $K=7$, the total running time is reduced by 75--97\% for all methods except label propagation.  The label propagation is extremely efficient, so it is hard to improve its run time for small graphs.

In terms of quality, we consider the modularity \cite{clauset2004finding, newman2004finding} in Fig.~\ref{figLFR2}. Observe that the modularity is nearly unchanged from 0.9 as $K$ increases.

To give another view of quality, we consider two cases for NMI \cite{danon2005comparing
} in Fig.~\ref{figLFR4} and Fig.~\ref{figLFR5}. 
Fig.~\ref{figLFR4} compares the results to ground truth (i.e., the true clusters). Again, the value is little changed for $K=5,6,7$, though here there is some small decrease.  If we did not know the ground truth, we could consider the NMI with respect to the original solution as shown in Fig.~\ref{figLFR5}. While a negative result is not damning, an NMI near 1 w.r.t the baseline result will indicate no major changes in the solution.
We use the definition of NMI in Ref. \cite{danon2005comparing}:
\begin{align}
I(A,B) = \frac{-2 \sum^{c_A}_{i=1}\sum^{c_B}_{j=1} N_{ij} \log(N_{ij}N/N_{i\cdot}N_{\cdot j})}{\sum^{c_A}_{i=1} N_{i\cdot} \log(N_{i\cdot}/N)+\sum^{c_B}_{j=1}N_{\cdot j} \log (N_{\cdot j}/N)}, 
\end{align}
where the number of real communities is denoted $c_A$, the number of found communities
is denoted $c_B$, $N_{ij}$ is the number of nodes in the real community $i$ that appear in the found community $j$, the sum over row $i$ of matrix $N_{ij}$ is denoted $N_{i\cdot}$, and the sum over column $j$ is denoted $N_{\cdot j}$.

\subsection{Results and Analysis --- Social Network Data}

\begin{figure*}
\centering
\begin{tabular}{cc}
 \subfloat[Graph size and details of relative run time for Louvain method \cite{blondel2008fast}]{\includegraphics[width=0.3\textwidth]{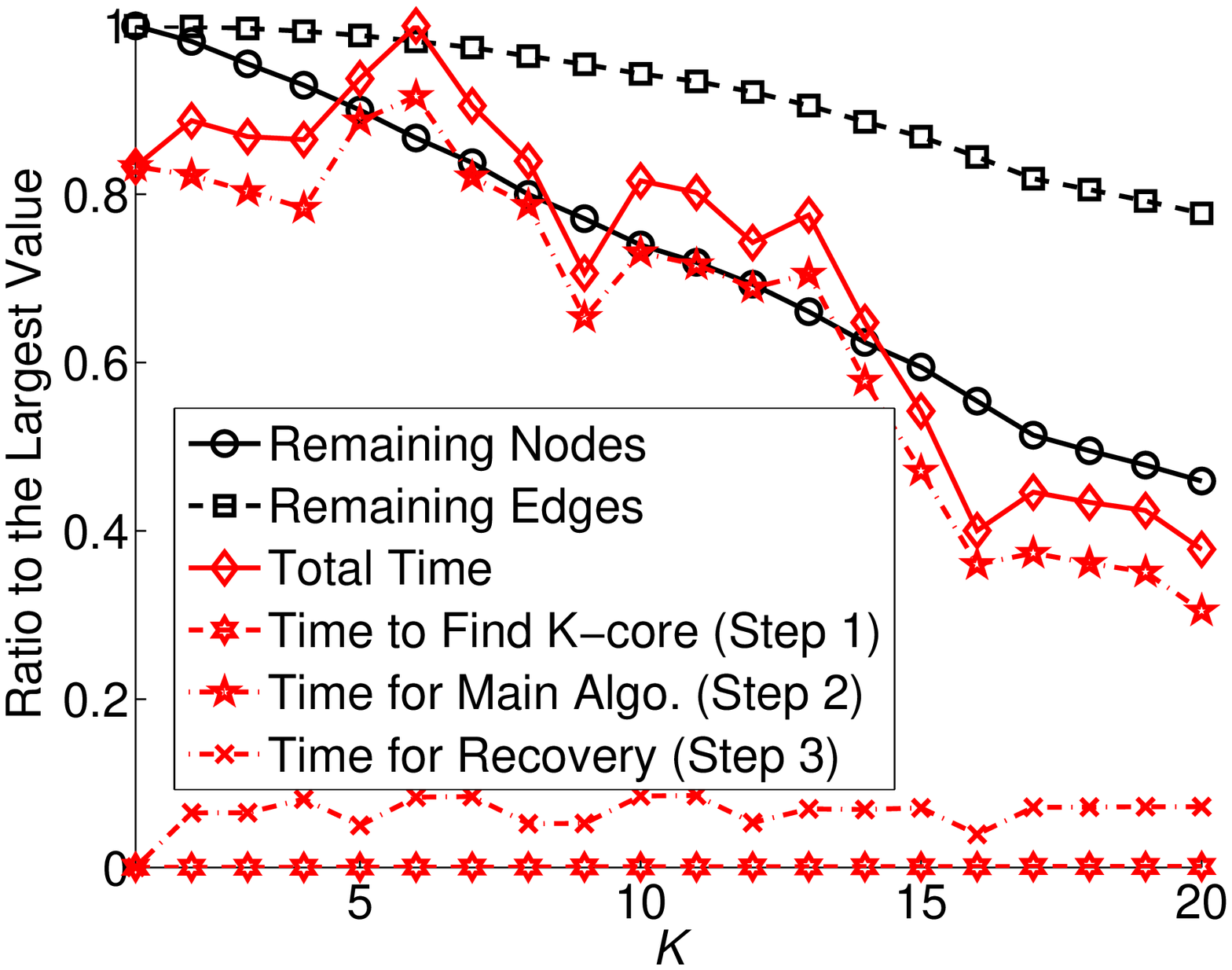}\label{figFacebook1}}
&
\subfloat[Relative run time]{\includegraphics[width=0.3\textwidth]{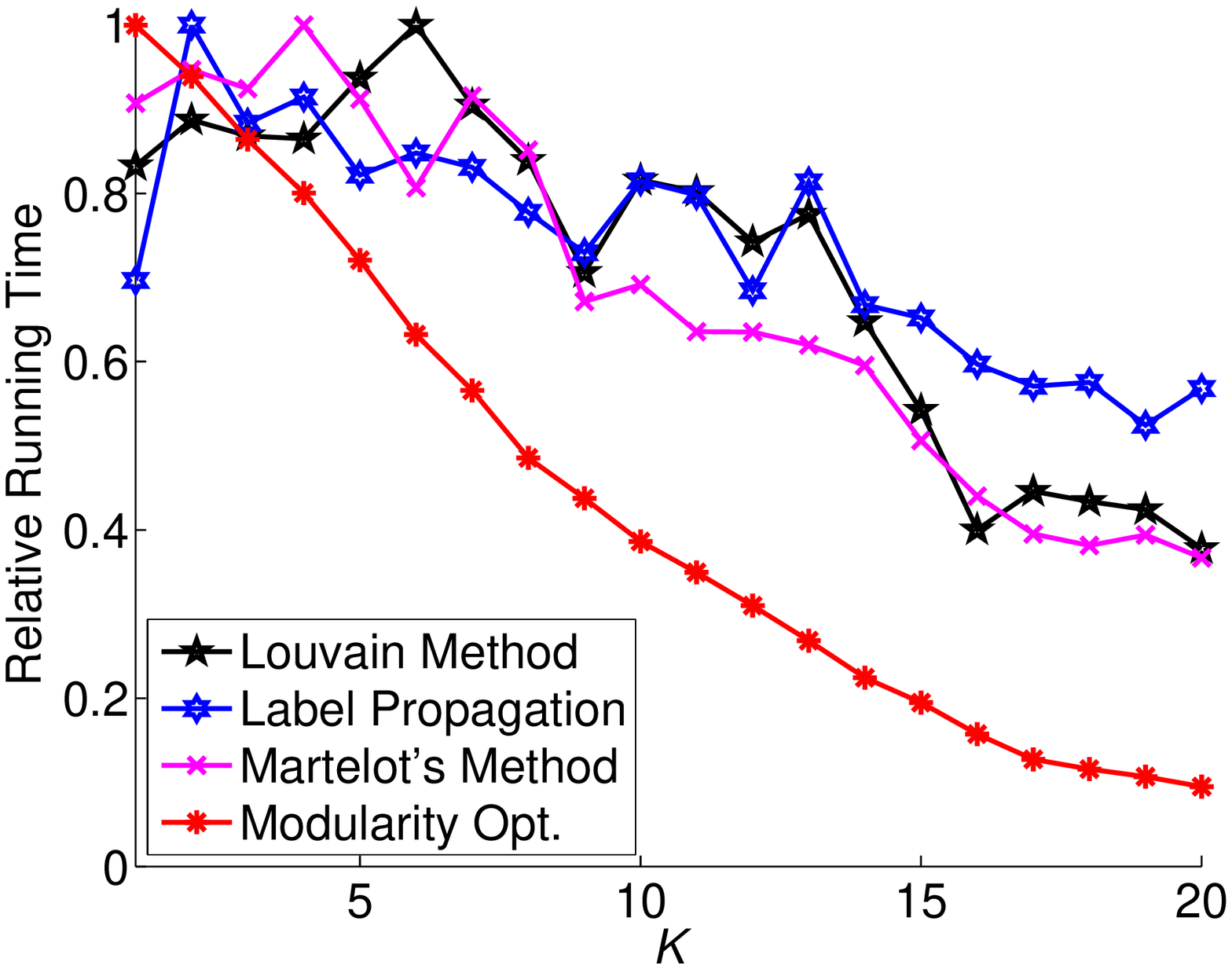}\label{figFacebook2}}
\end{tabular}
\begin{tabular}{ccc}
\subfloat[Modularity]{\includegraphics[width=0.3\textwidth]{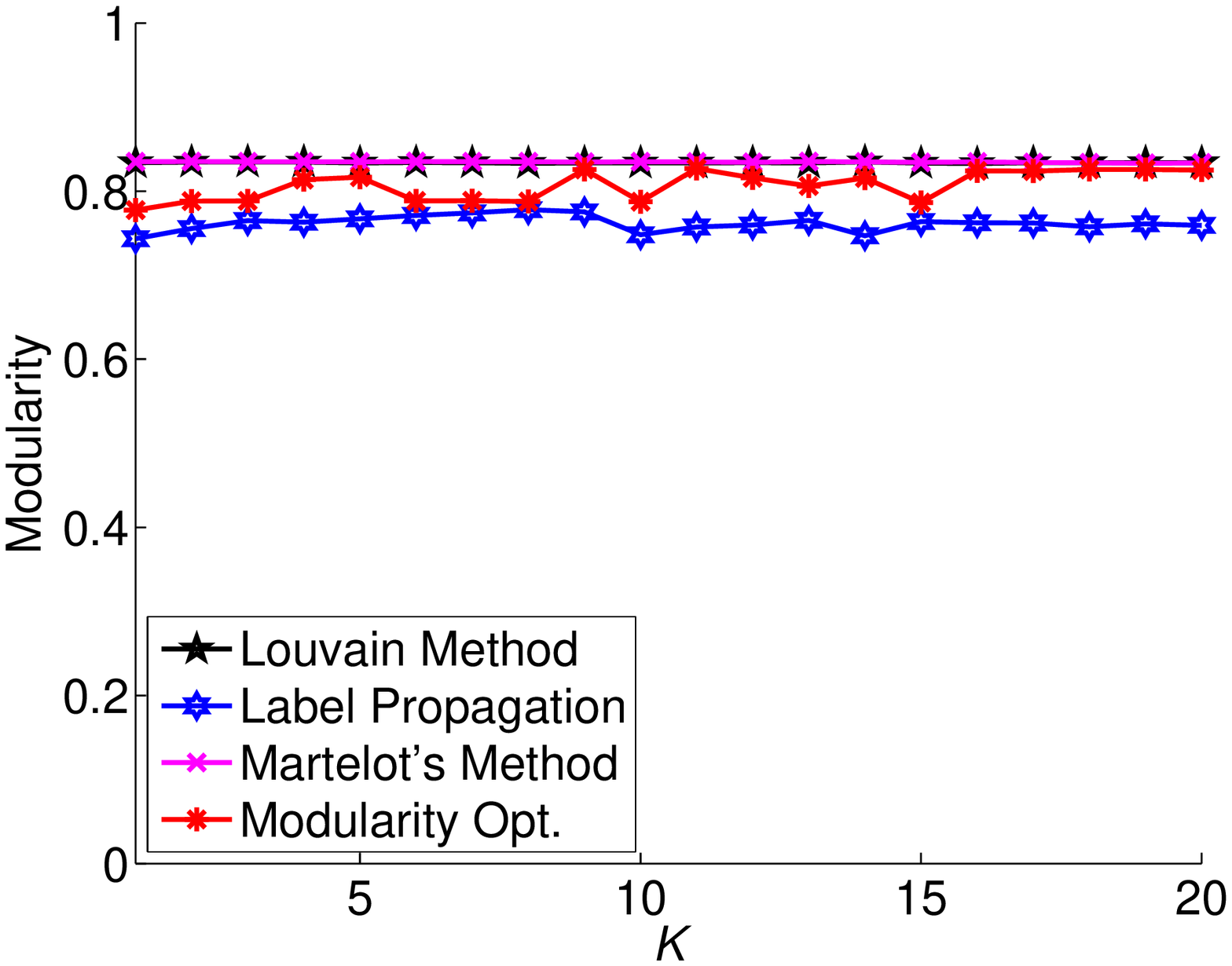}\label{figFacebook3}}
&
\subfloat[NMI w.r.t.\@  ground truth]{\includegraphics[width=0.3\textwidth]{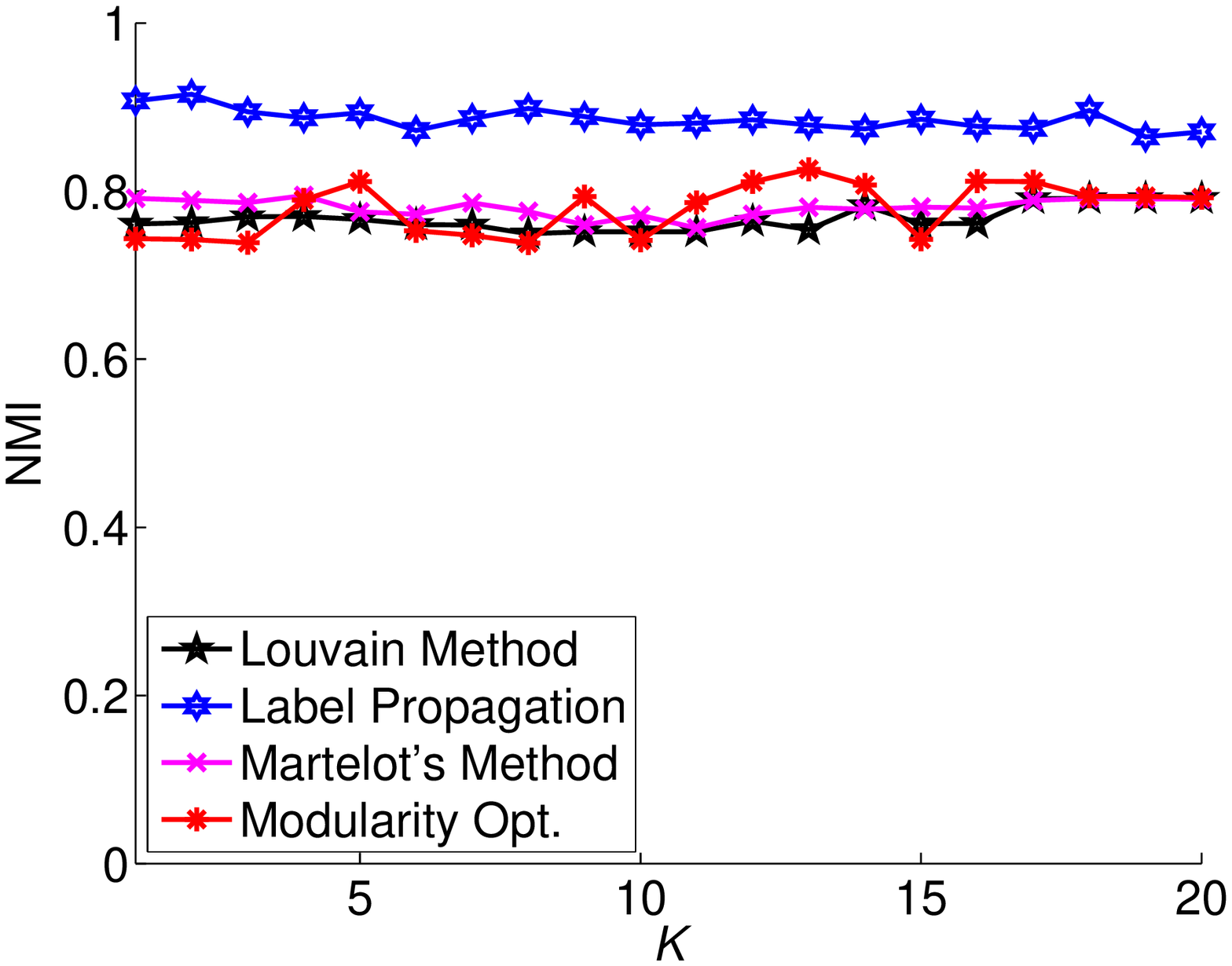}\label{figFacebook4}}
&
\subfloat[NMI w.r.t.\@  baseline result]{\includegraphics[width=0.3\textwidth]{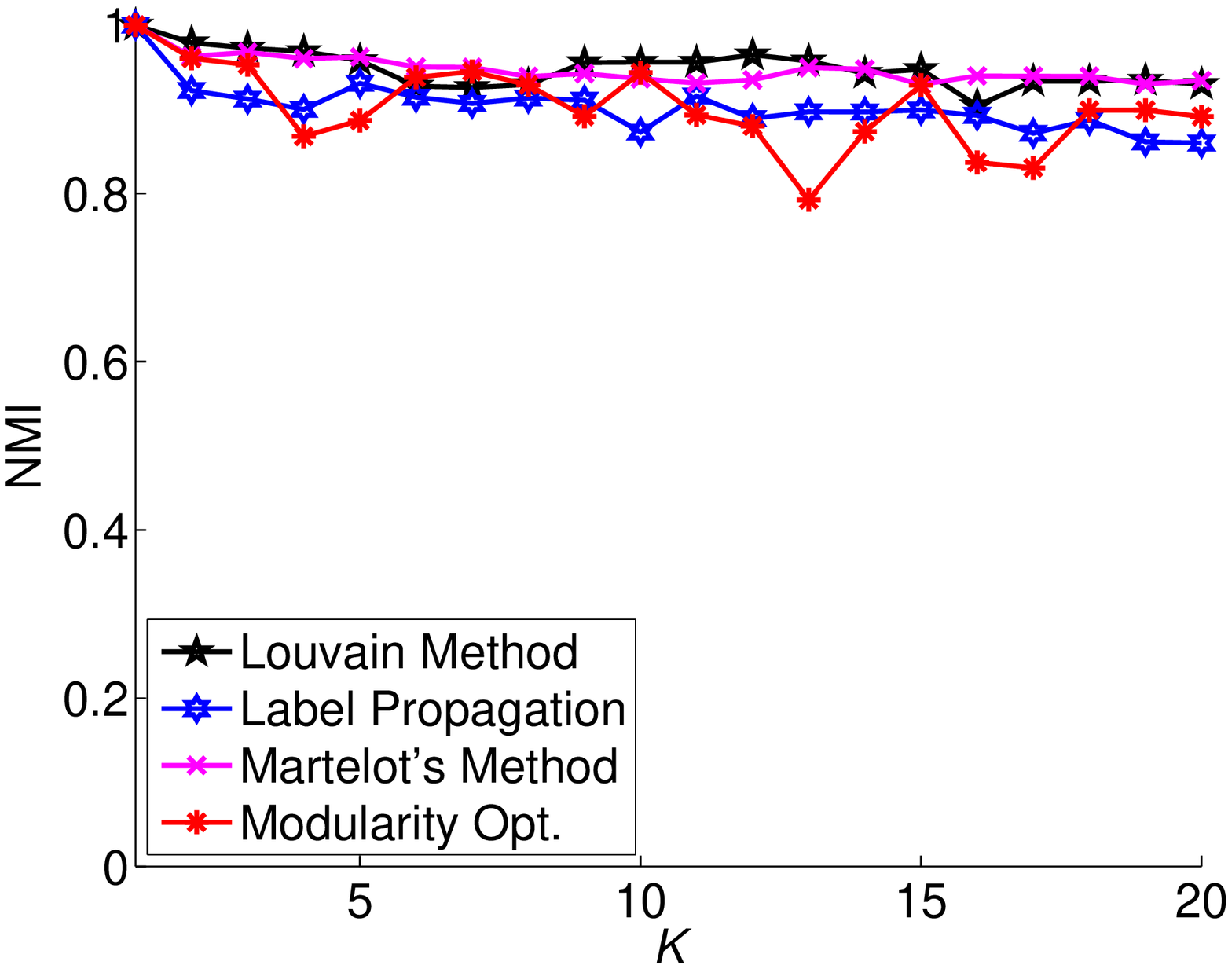}\label{figFacebook5}}
\end{tabular}
\caption{Results of $K$-core acceleration on ego-Facebook}
\label{figFacebook}
\end{figure*}



Fig.~\ref{figFacebook} shows analogous results to Fig.~\ref{figLFR}, except that it is on the ego-Facebook graph, in which we consider the ego networks constituting the graph as the ground-truth communities.\footnote{We have access to overlapping community information. If a node is in multiple communities, we assign it to the community to which is has the most connections, breaking ties randomly.}

In Fig.~\ref{figFacebook1}, the highest runtime for the Louvain method is actually for $K=7$, approximately 25\% more expensive than for the full graph. However, as $K$ continues to increase, the runtime reduces to less than 50\% of the runtime for the full graph. Once again, Step 2 requires the majority of the computation time but eventually reduces for the smaller $K$-cores.

In Fig.~\ref{figFacebook2}, we see that all but one of the methods have some fluctuation in runtimes for lower values of $K$. The reason is that, as $K$ increases, the number of iterations may change even though the cost per iteration is less. For example, the number of iterations in  the Louvain method \cite{blondel2008fast} increases from 9 to 11 when $K$ is increased from 1 to 5. Nevertheless, the total running time will decrease eventually. Therefore, although the running time fluctuates within a small range, the general trend is decreasing as $K$ grows.

In terms of quality, the Figs.~\ref{figFacebook3}--\ref{figFacebook5} show no degradation.  The NMI w.r.t. the baseline in Fig. ~\ref{figFacebook5} shows that there is little change w.r.t.\@ the original solution as well. We also note that label propagation has the lowest score with respect to modularity but the highest score w.r.t.\@ NMI compared to ground truth.

\subsection{Results and Analysis --- Large-scale networks}

\begin{figure*}[!ht]
\centering
\begin{tabular}{cc}
\subfloat[Modularity for com-Youtube]{\includegraphics[width=0.3\textwidth]{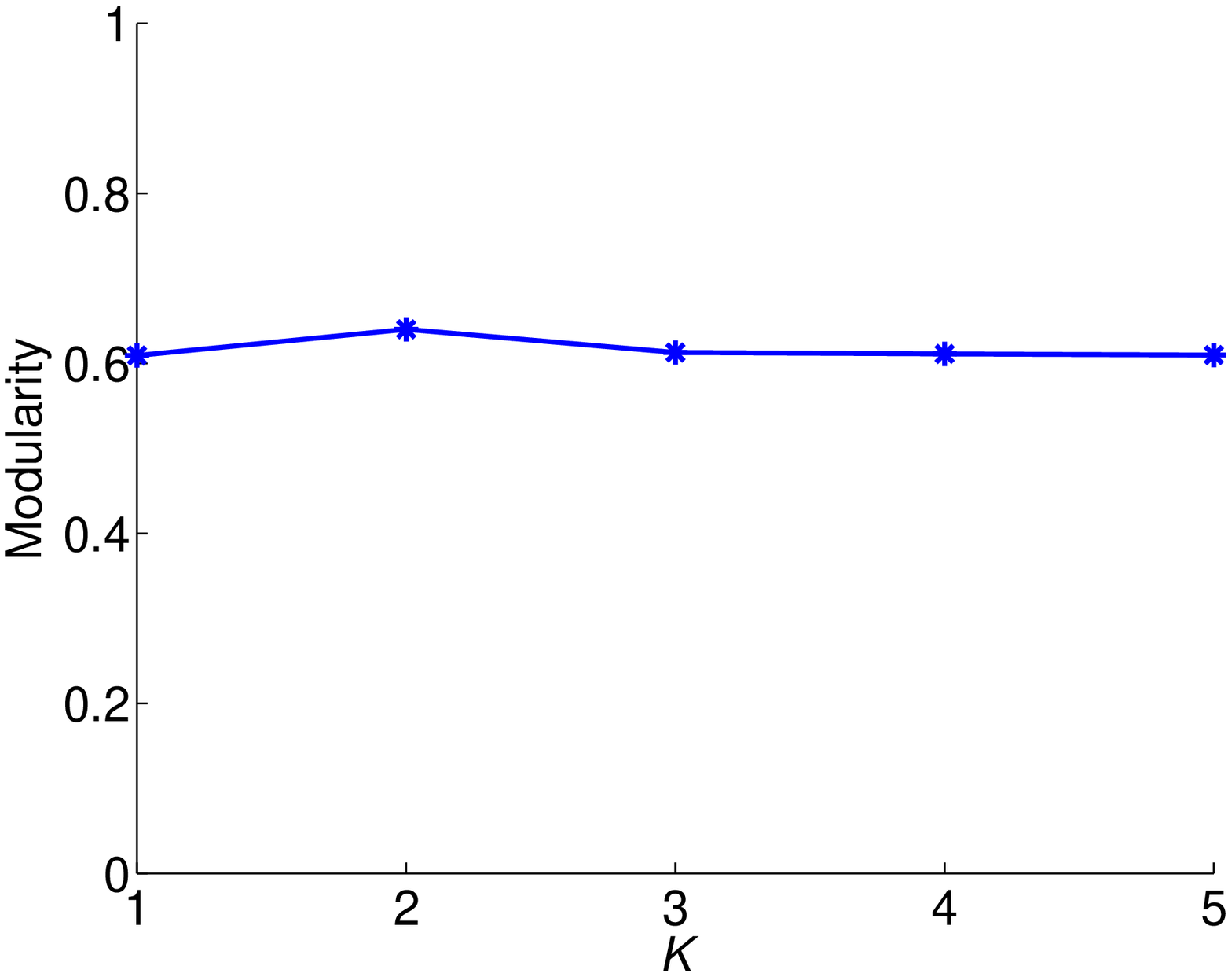}}
&
\subfloat[Graph size and details of relative run  time for label propagation on com-Youtube]{\includegraphics[width=0.3\textwidth]{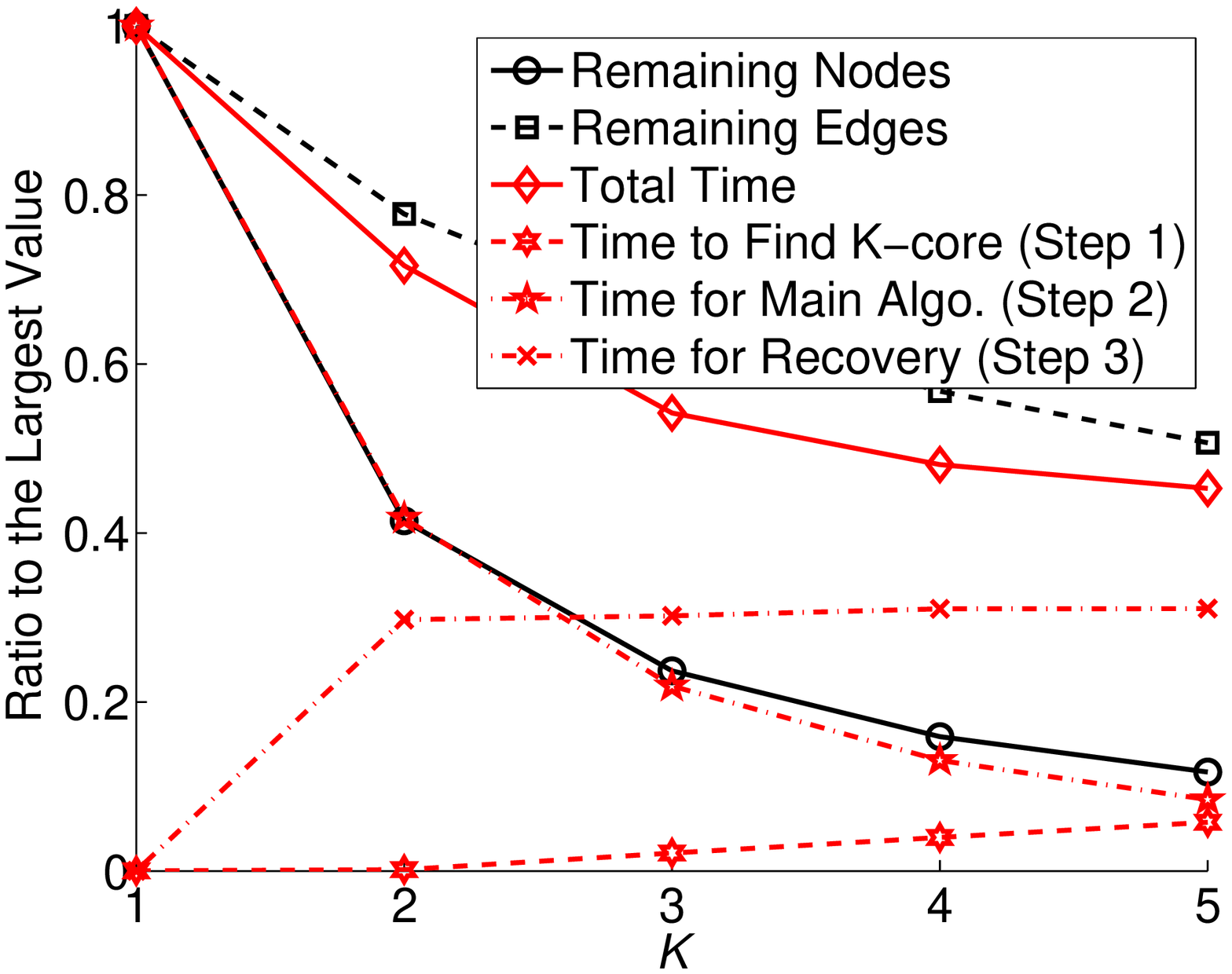}}
\\
\subfloat[Modularity for com-Amazon]{\includegraphics[width=0.3\textwidth]{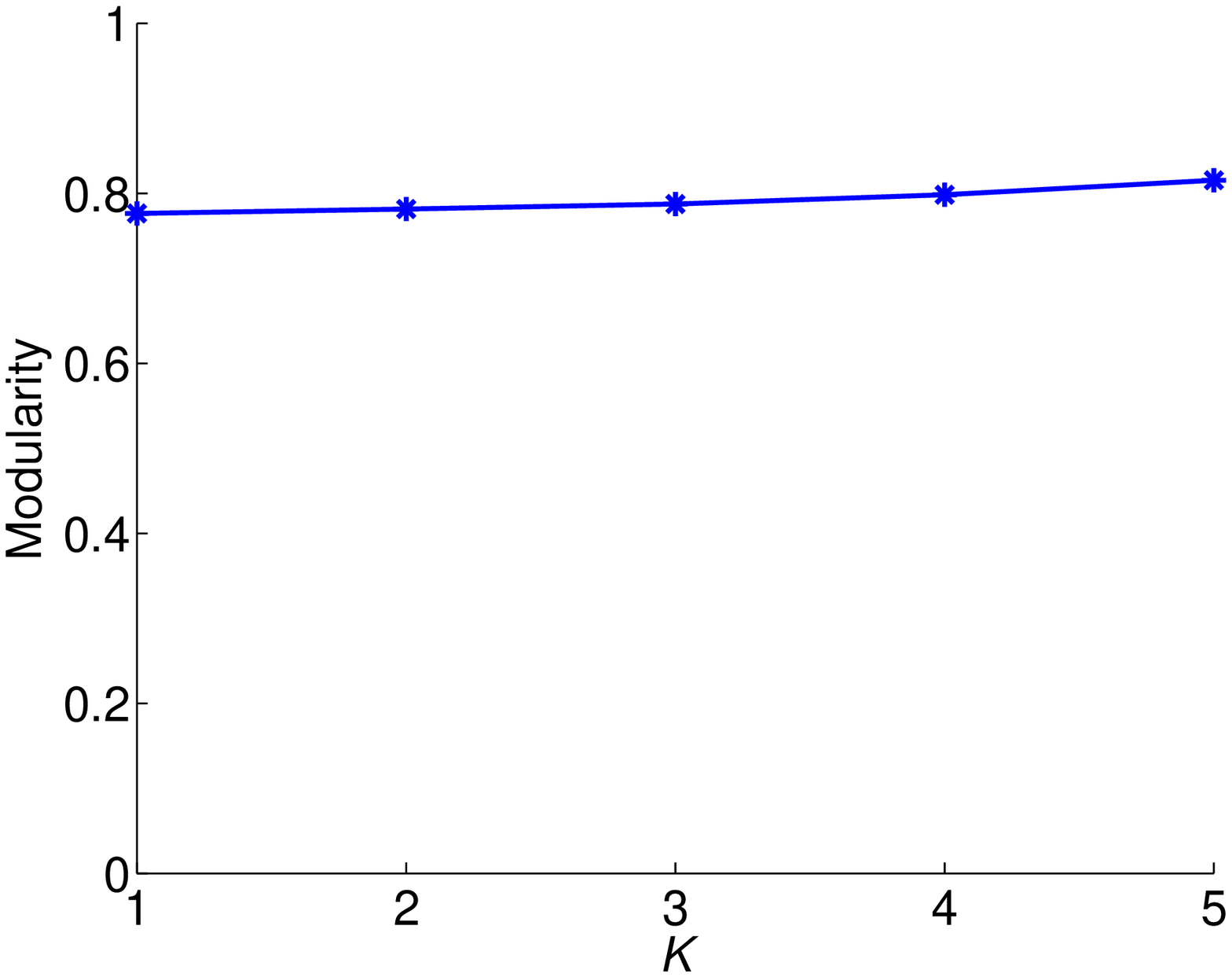}}
&
\subfloat[Graph size and details of relative run  time for label propagation on  com-Amazon]{\includegraphics[width=0.3\textwidth]{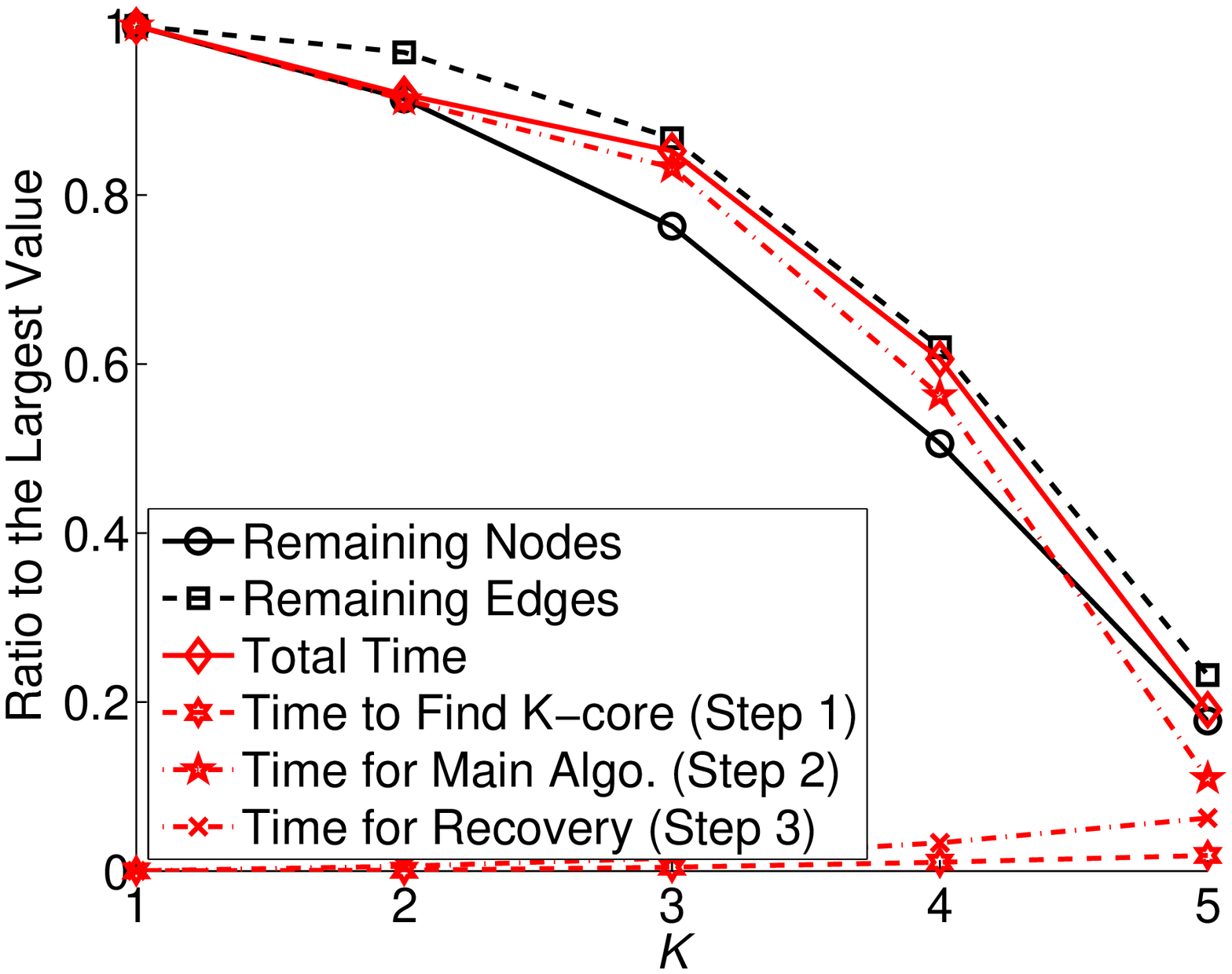}}
\end{tabular}
\caption{Results of $K$-core acceleration on com-Youtube and com-Amazon for Community Detection}
\label{figDblpAmazon}
\end{figure*}

Fig.~\ref{figDblpAmazon} demonstrates how our $K$-core framework accelerates the (linear-time) label propagation algorithm \cite{raghavan2007near} on com-Youtube and com-Amazon data.
Since these are large graphs, we skip the modularity optimization (i.e., the second stage algorithm) in Step (3). Any unlabeled nodes are assigned to singleton communities.
%

The plots on the left of Fig.~\ref{figDblpAmazon} show that the quality (based on  modularity) is unchanged as $K$ increases.
The plots on the right of Fig.~\ref{figDblpAmazon} show details on the method. In these cases, the label propagation algorithm used in Step 2 is so inexpensive that the cost of Step 3 (even without the modularity optimization) eventually dominates the total time. Nevertheless, the overall run time is significantly reduced even when the original algorithm is only linear!


{ 
Louvain method \cite{blondel2008fast} is also a linear time algorithm
, but the average running time for each node is much longer than label propagation.  Thus, we only run those algorithms with $K \geq 3$ on the com-Amazon network. By Fig.~\ref{figDblpAmazon}, the ratio of remaining nodes drops from $76\%$ in 3-core to 50\% in 4-core, 17\% in 5-core, and 0.2\% in 6-core. In Table  \ref{tableAmazon}, the  running time column contains the running time in seconds, and those in the bracket behind are the relative running times with respect to 3-core. We show that by our approach, the modularity computed through both algorithms does not change much from 3-core to 5-core, but the running times have significant reduction. At 6-core, because the remaining nodes are too few, the modularity is no longer a satisfactory.

\begin {table}
\footnotesize
\caption {Algorithms Using $K$-core on com-Amazon network} \label{tableAmazon}
\begin{center}
\begin{tabular}{l*{3}{c}r}
  & K & Modularity  & Time (Relative to 3-core) 
\\
\hline
& 3 & 0.923 & 6.3e4  (1.00)\\
	& 4 & 0.917 & 2.6e4 (0.41) \\
	& 5 & 0.890&  3.2e3  (0.05)\\
	& 6 & 0.788& 5.5e2 (0.01)\\
\end{tabular}
\end{center}
\end{table}
}

\subsection{Results and Analysis --- Multiple real-world networks}

\newcommand{\figwidthStepAll}{0.32\textwidth}

\begin{figure*}
\centering
\begin{tabular}{ccc}
\subfloat[ca-GrQc]{\includegraphics[width=\figwidthStepAll]{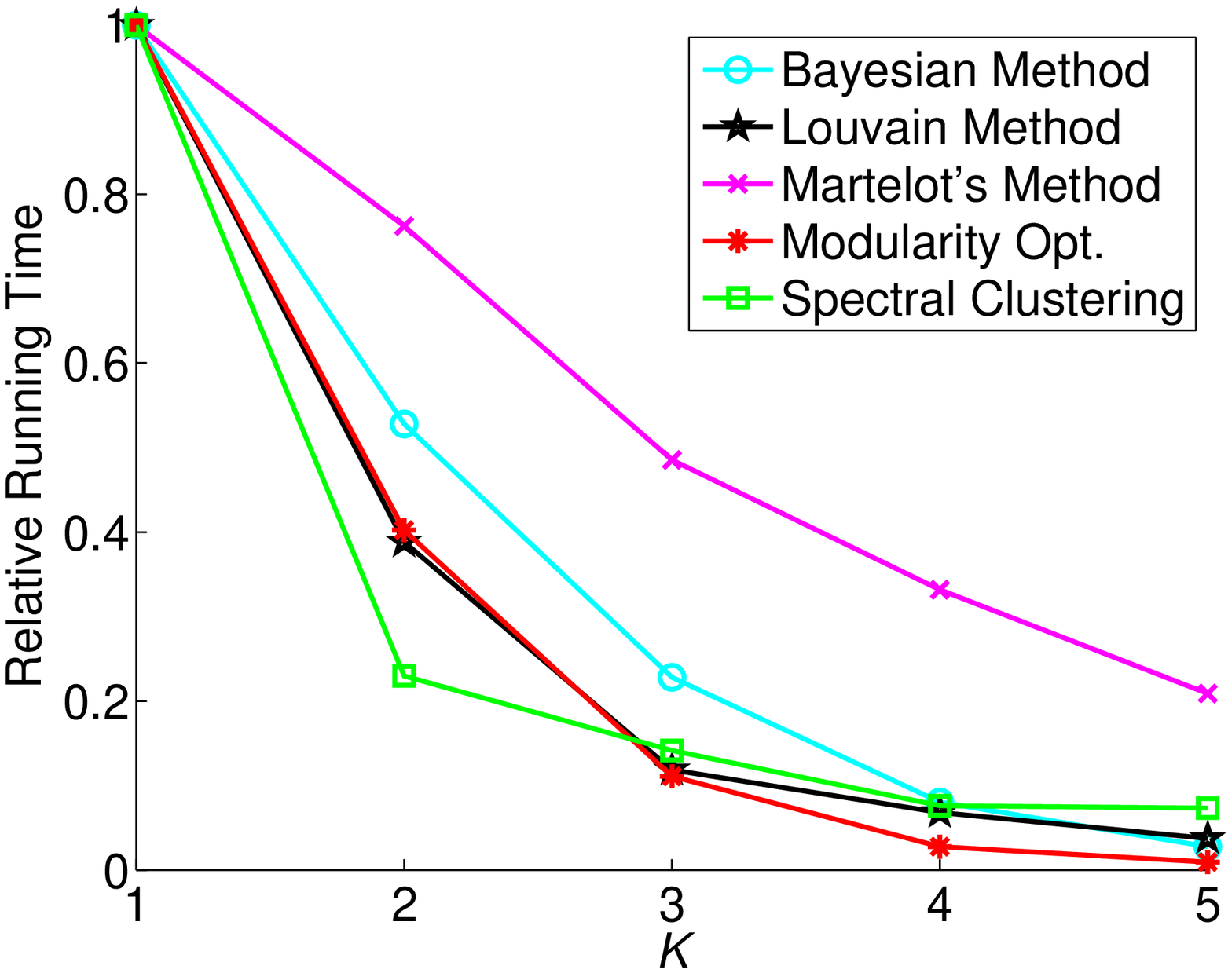}\label{figGrQcModTime}}  &
\subfloat[ca-HepPh]{\includegraphics[width=\figwidthStepAll]{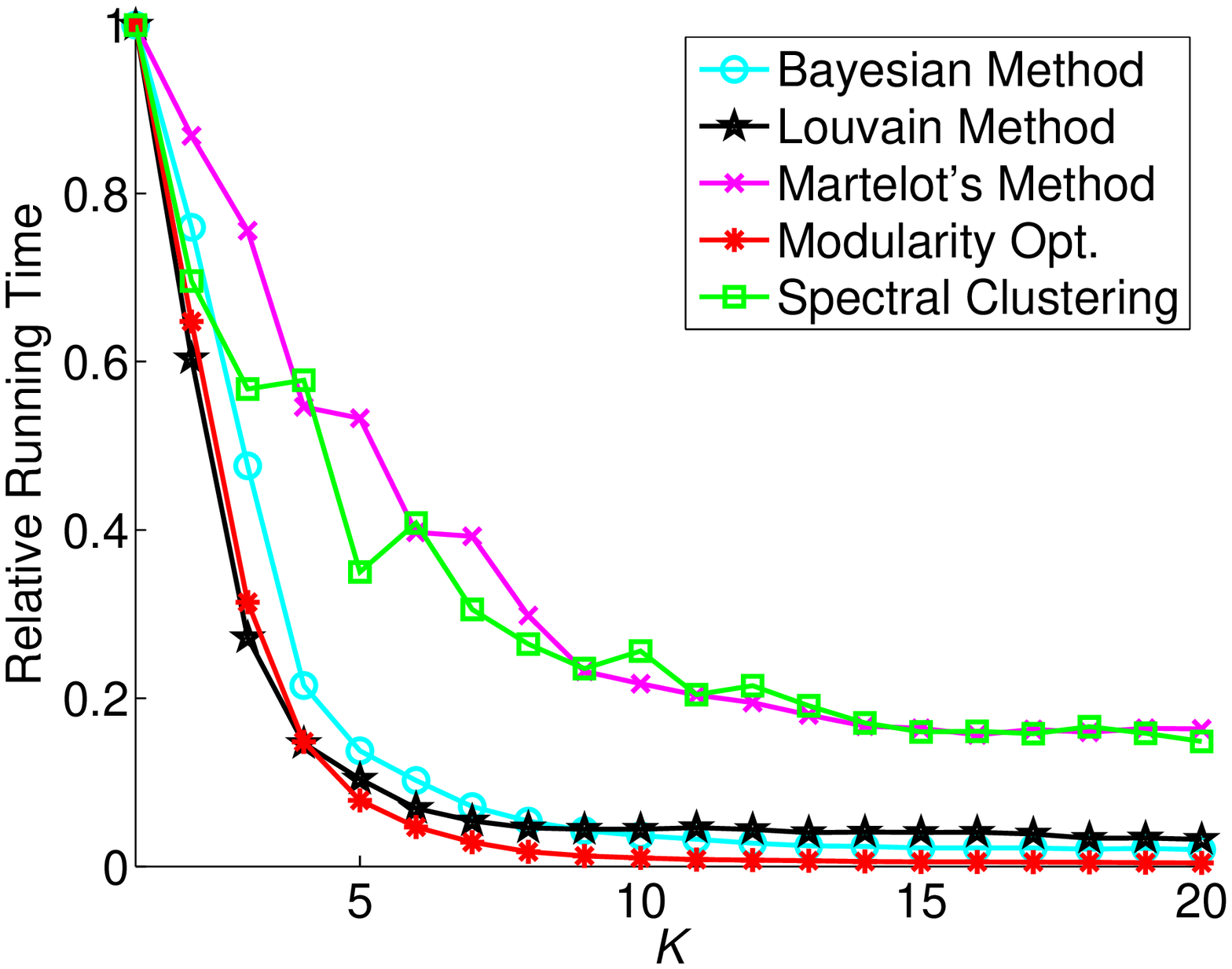}} &
\subfloat[ca-HepTh]{\includegraphics[width=\figwidthStepAll]{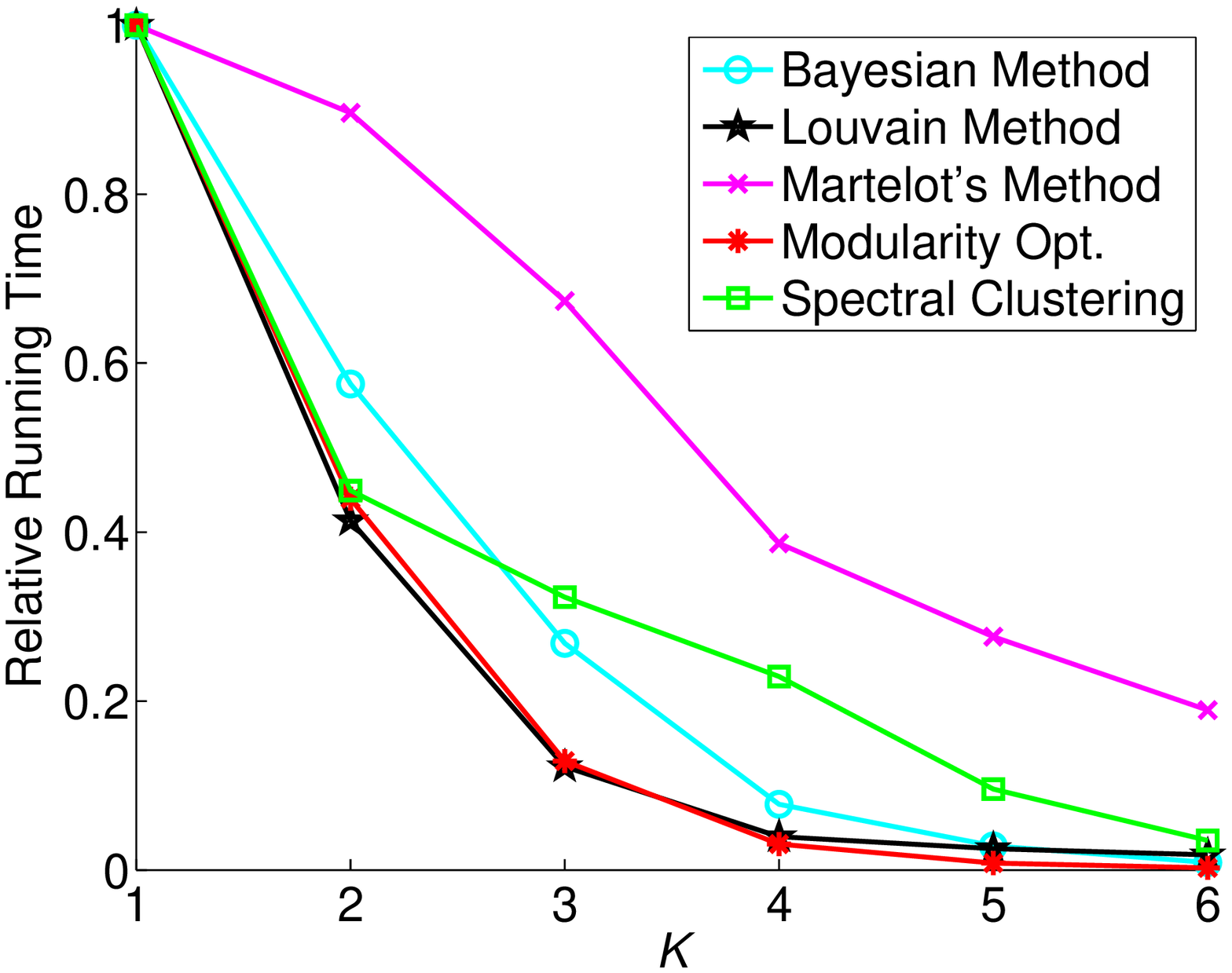}\label{allTime3}}
\\
\subfloat[ca-CondMat]{\includegraphics[width=\figwidthStepAll]{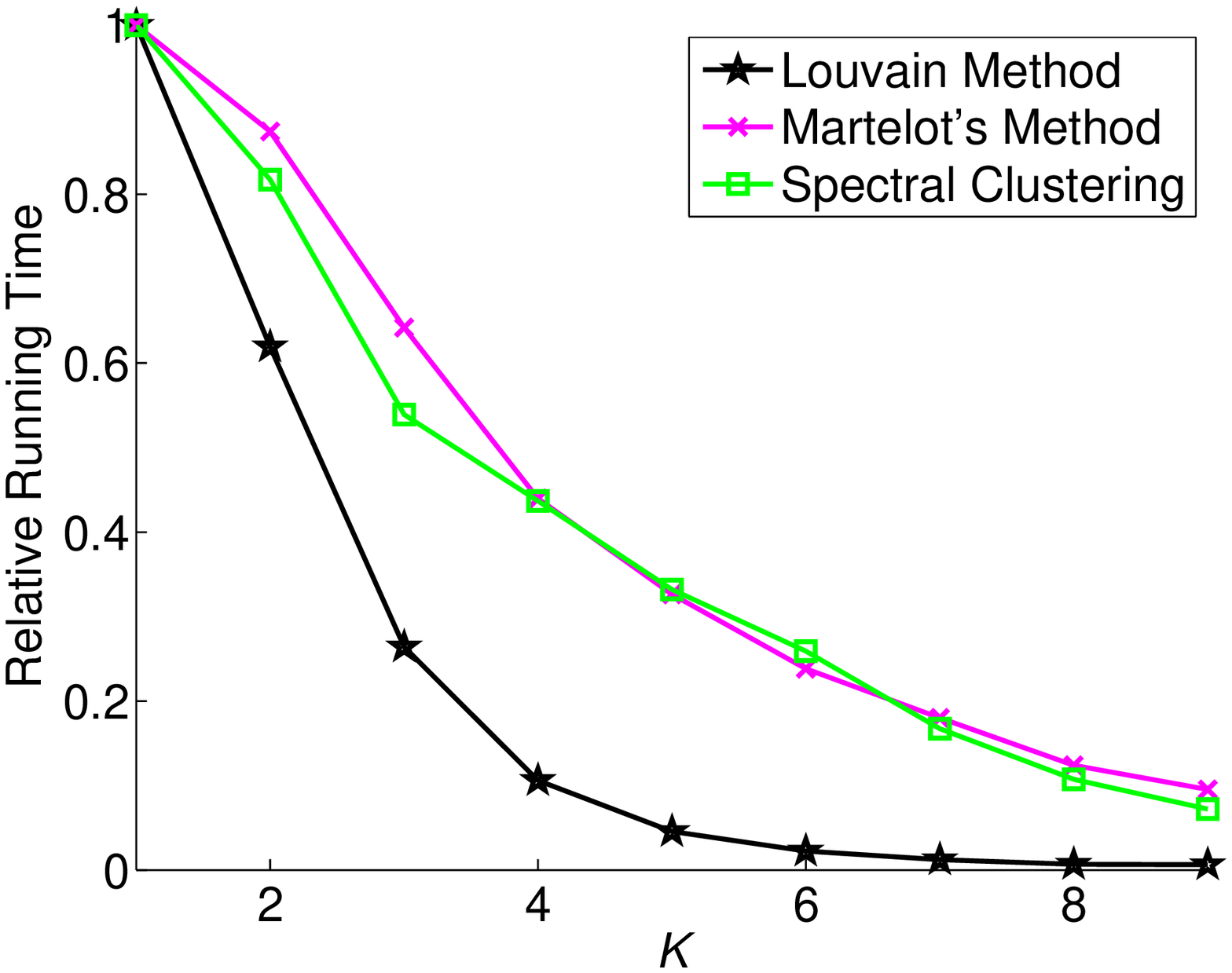}\label{allTime4}} &
\subfloat[ca-AstroPh]{\includegraphics[width=\figwidthStepAll]{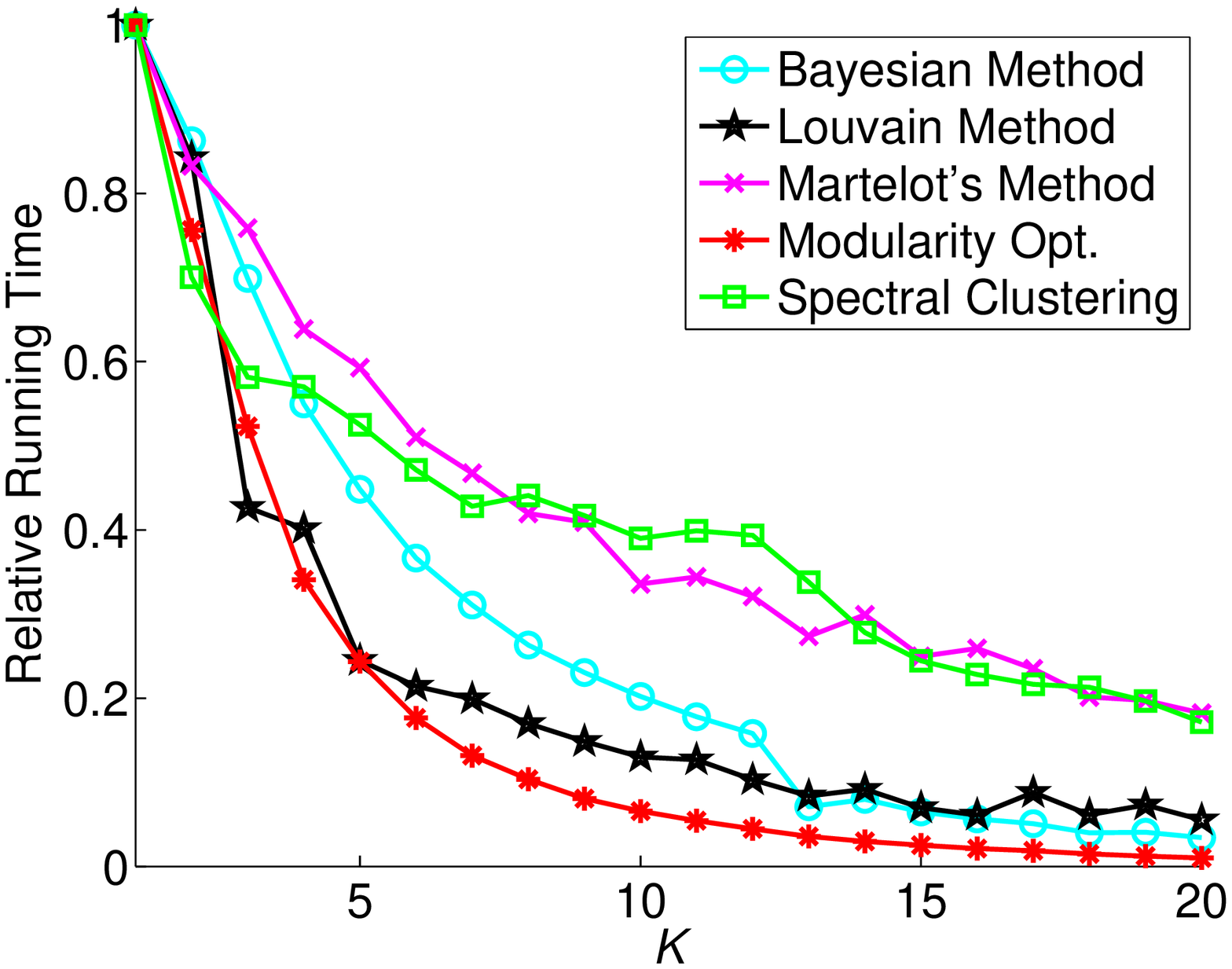}}
&
\subfloat[Email-Enron]{\includegraphics[width=\figwidthStepAll]{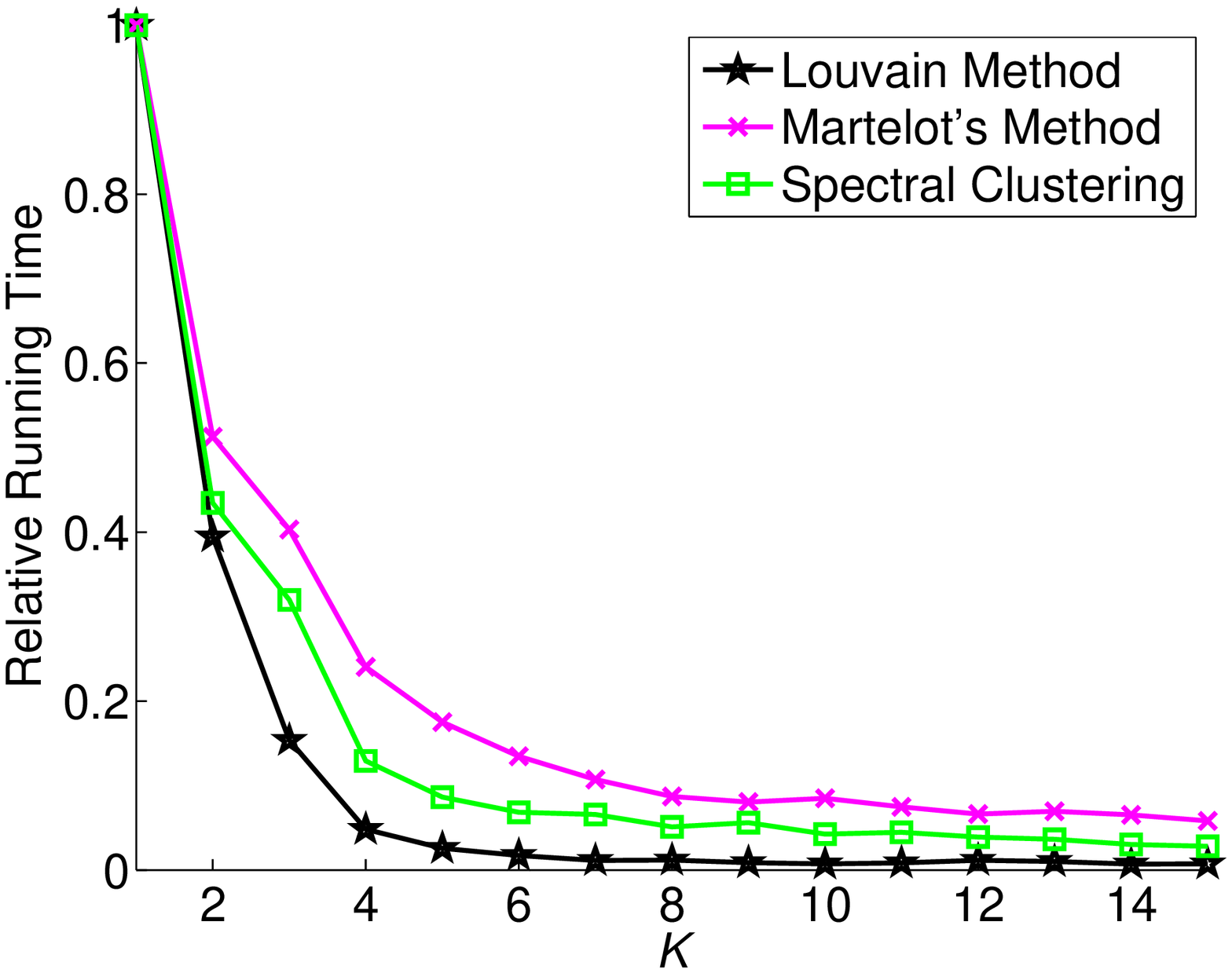}} \\
\subfloat[oregon1\_010331]{\includegraphics[width=\figwidthStepAll]{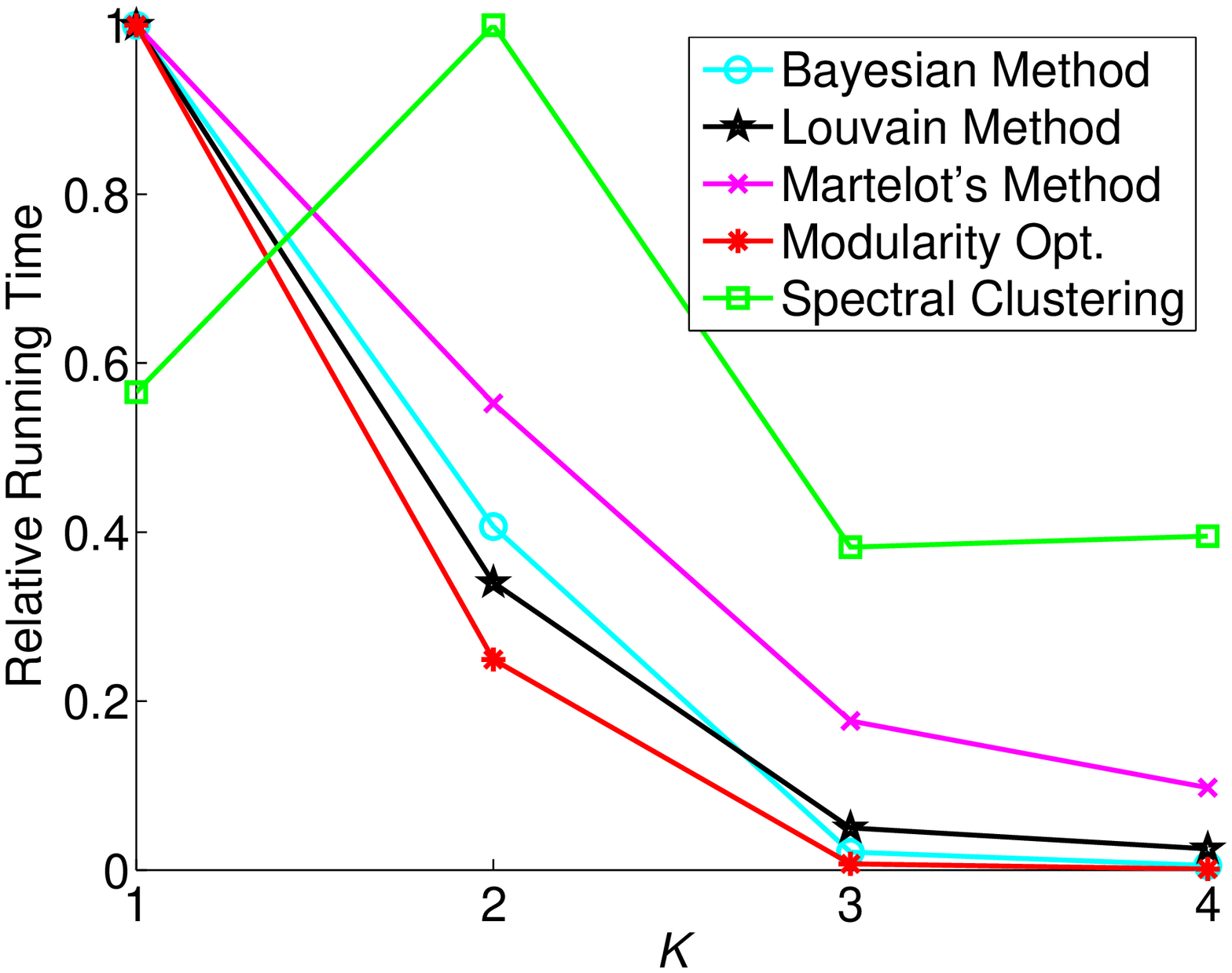}}
&
\subfloat[oregon1\_010421]{\includegraphics[width=\figwidthStepAll]{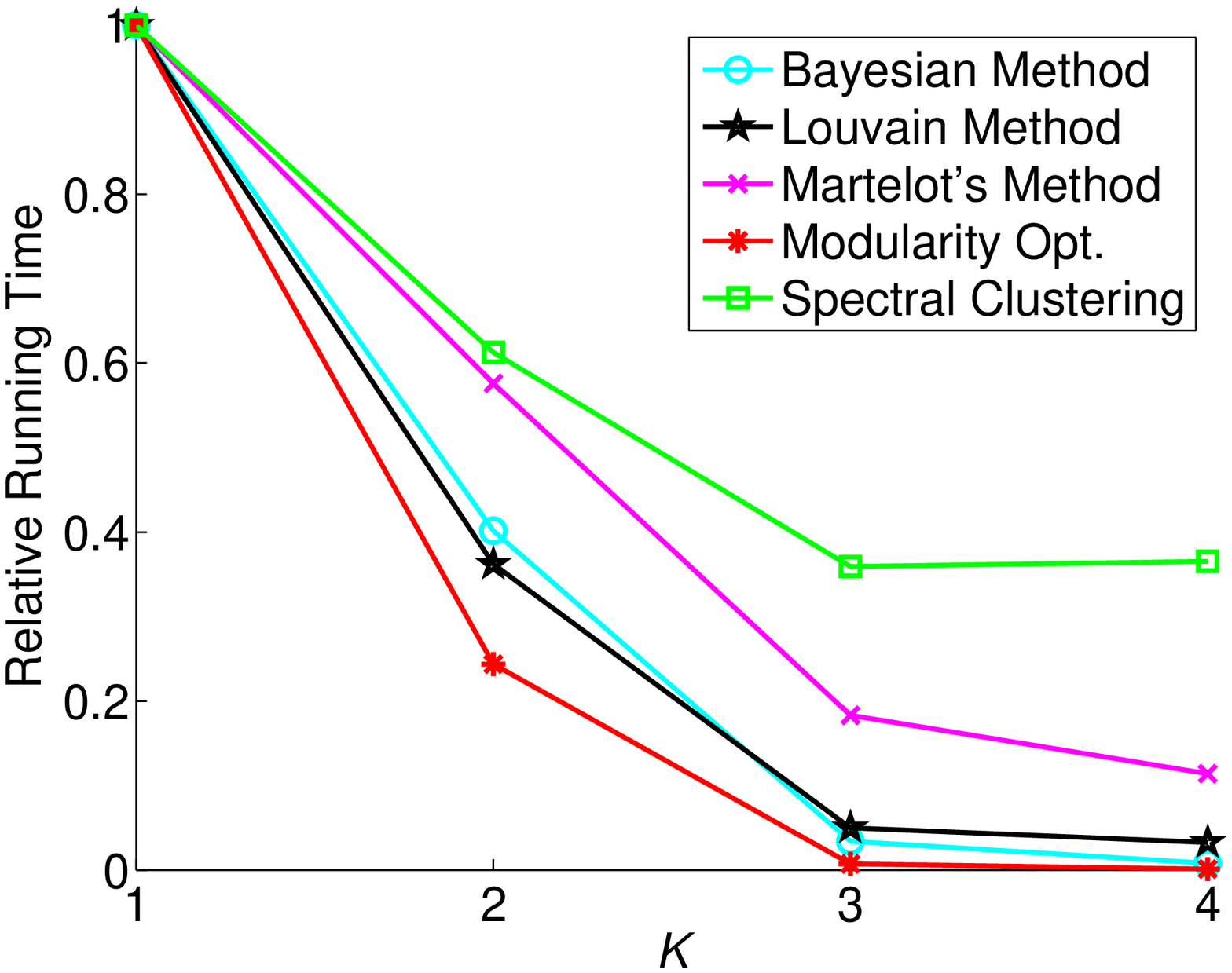}} &
\subfloat[oregon1\_010428]{\includegraphics[width=\figwidthStepAll]{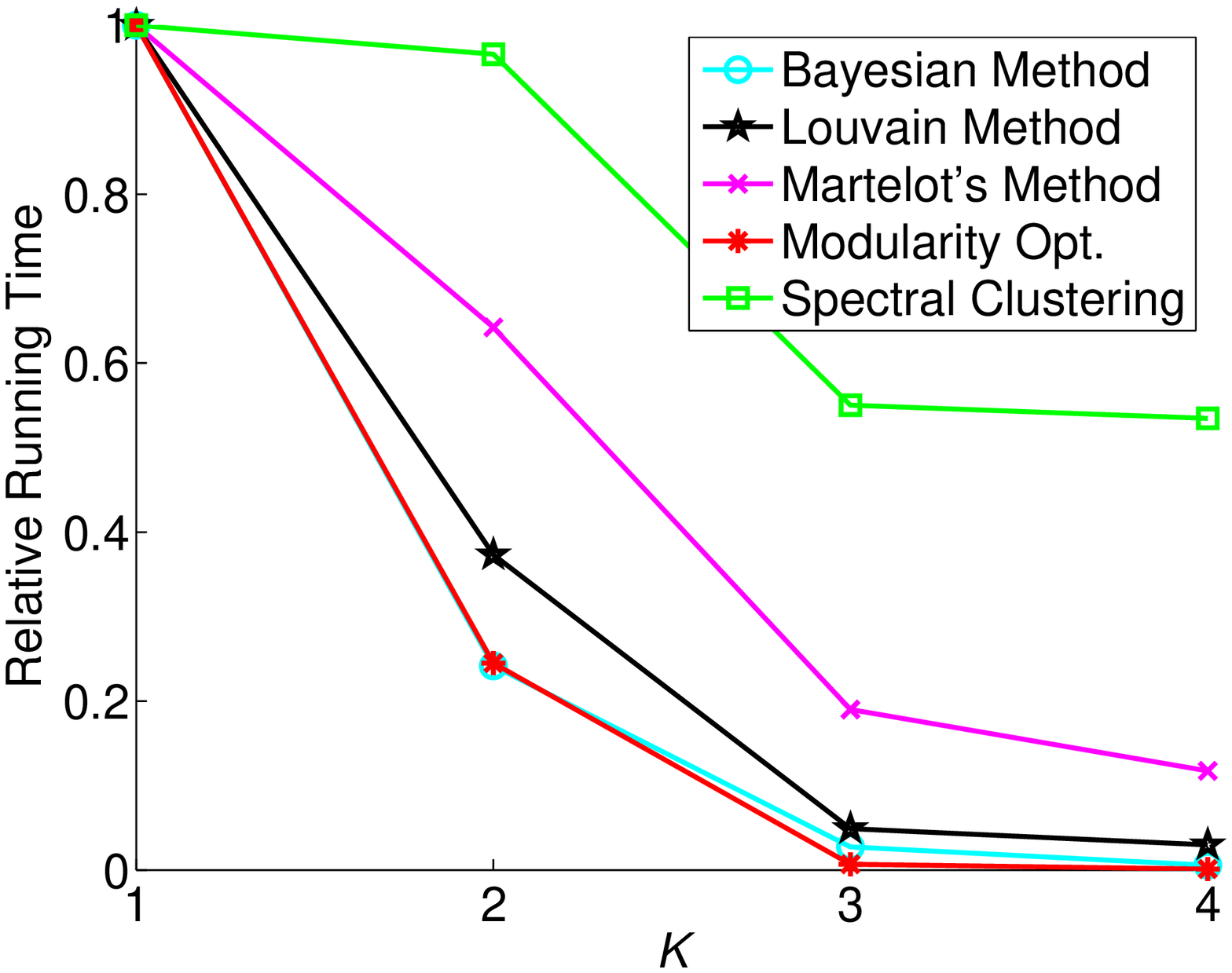}} \\
\end{tabular}
\caption{Relative run time over different $K$'s} \label{allTime}
\end{figure*}

\begin{figure*}
\centering
\begin{tabular}{ccc}
\subfloat[ca-GrQc]{\includegraphics[width=\figwidthStepAll]{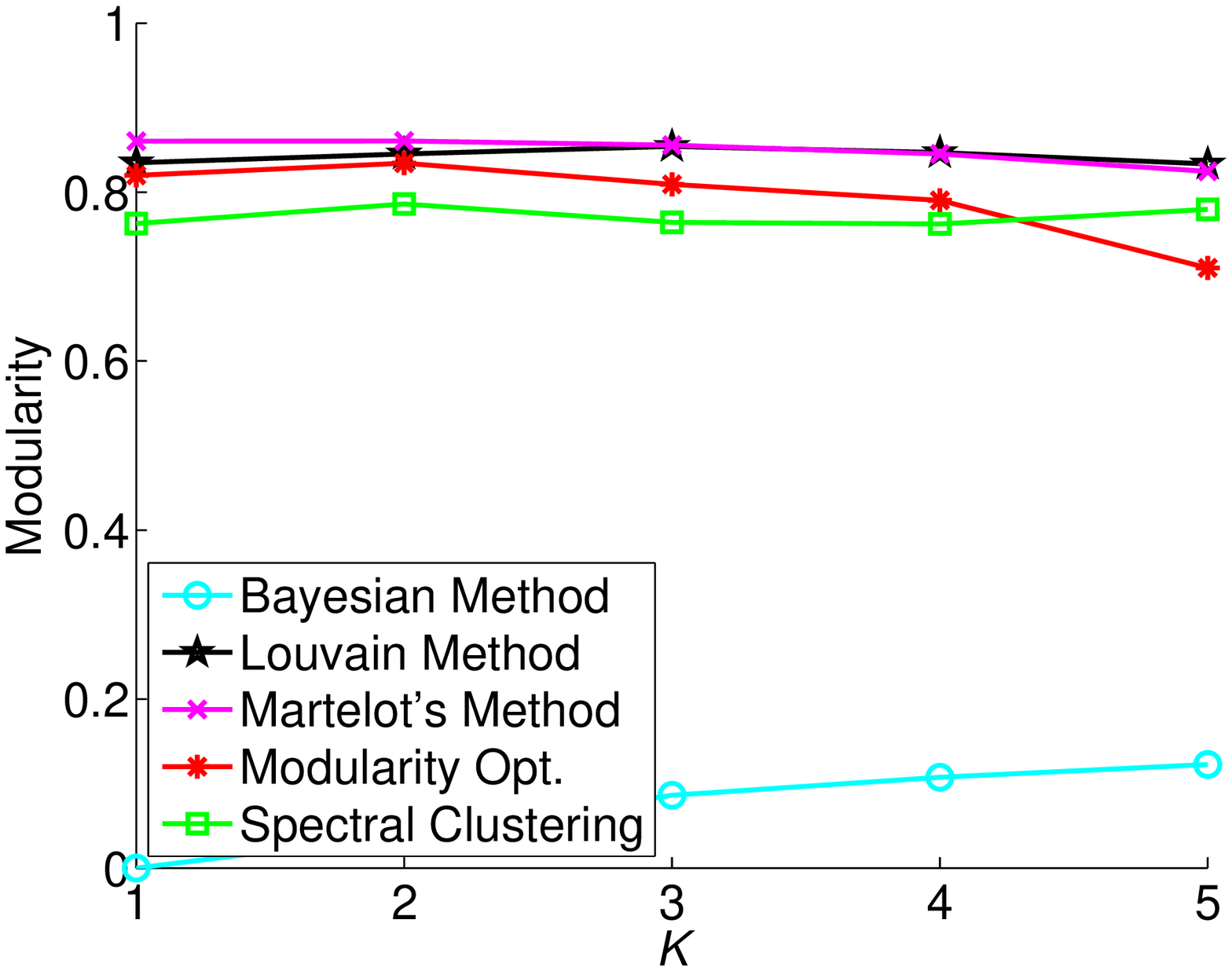}\label{figGrQcModMod}}
&
\subfloat[ca-HepPh]{\includegraphics[width=\figwidthStepAll]{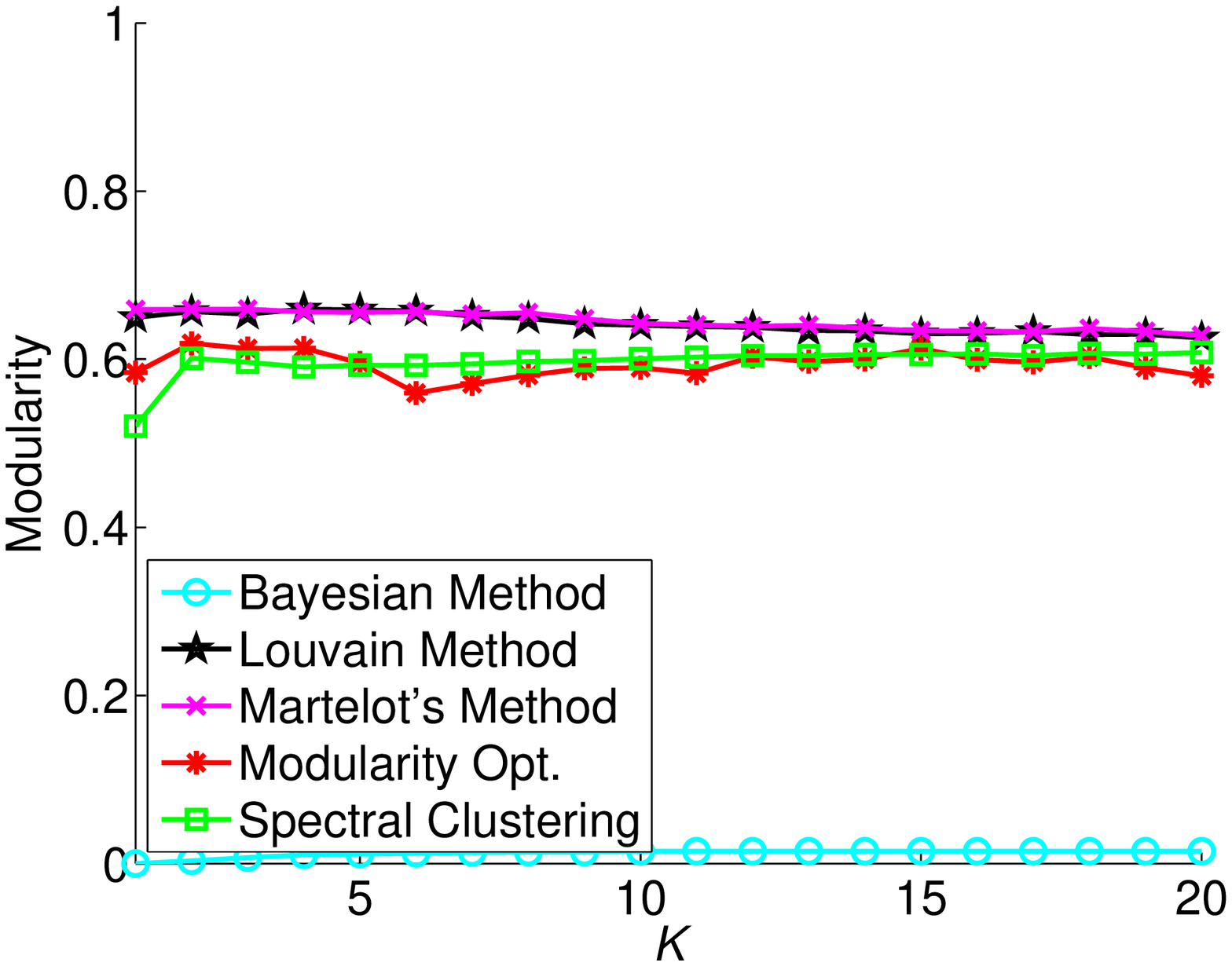}} &
\subfloat[ca-HepTh]{\includegraphics[width=\figwidthStepAll]{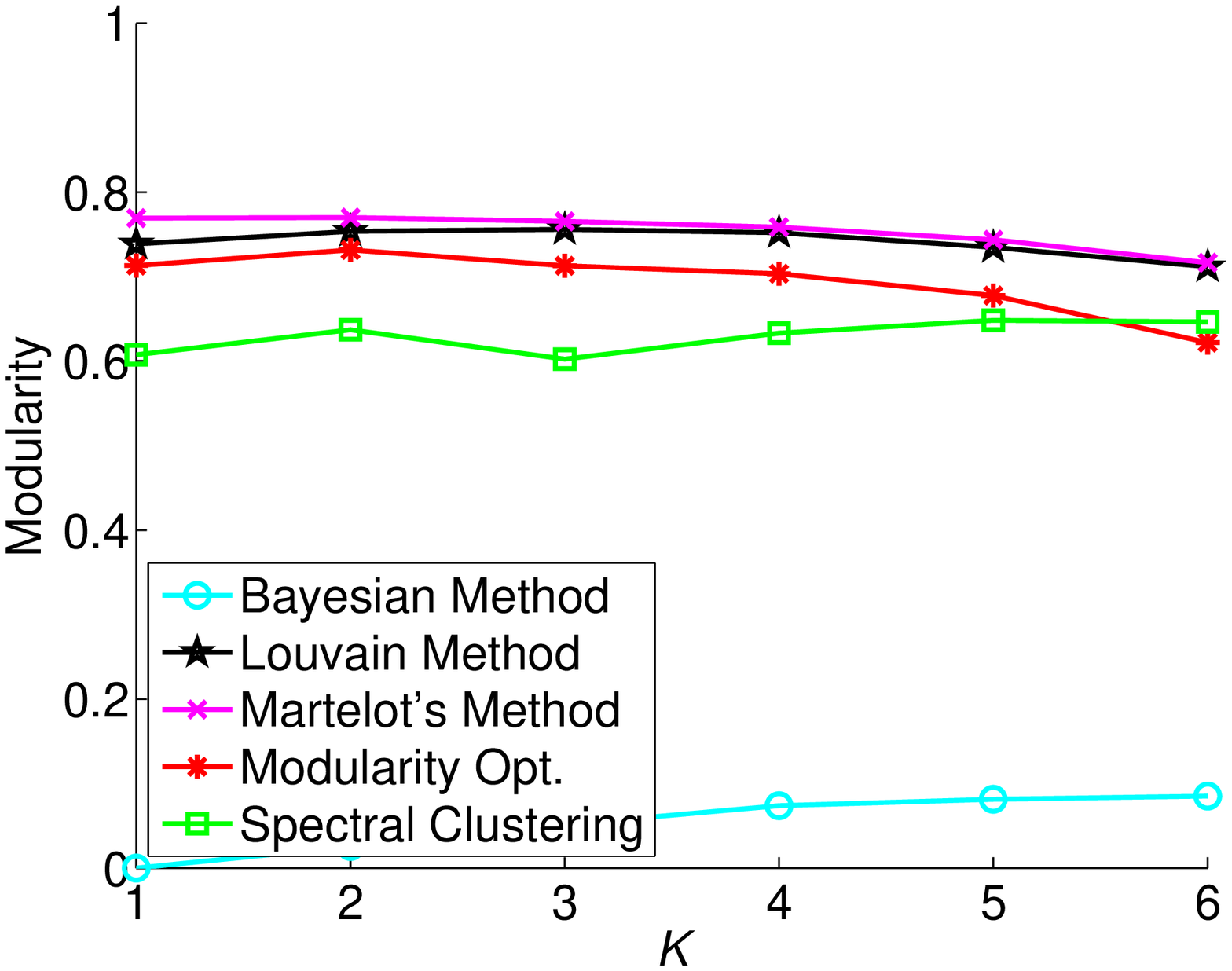}}
\\
\subfloat[ca-CondMat]{\includegraphics[width=\figwidthStepAll]{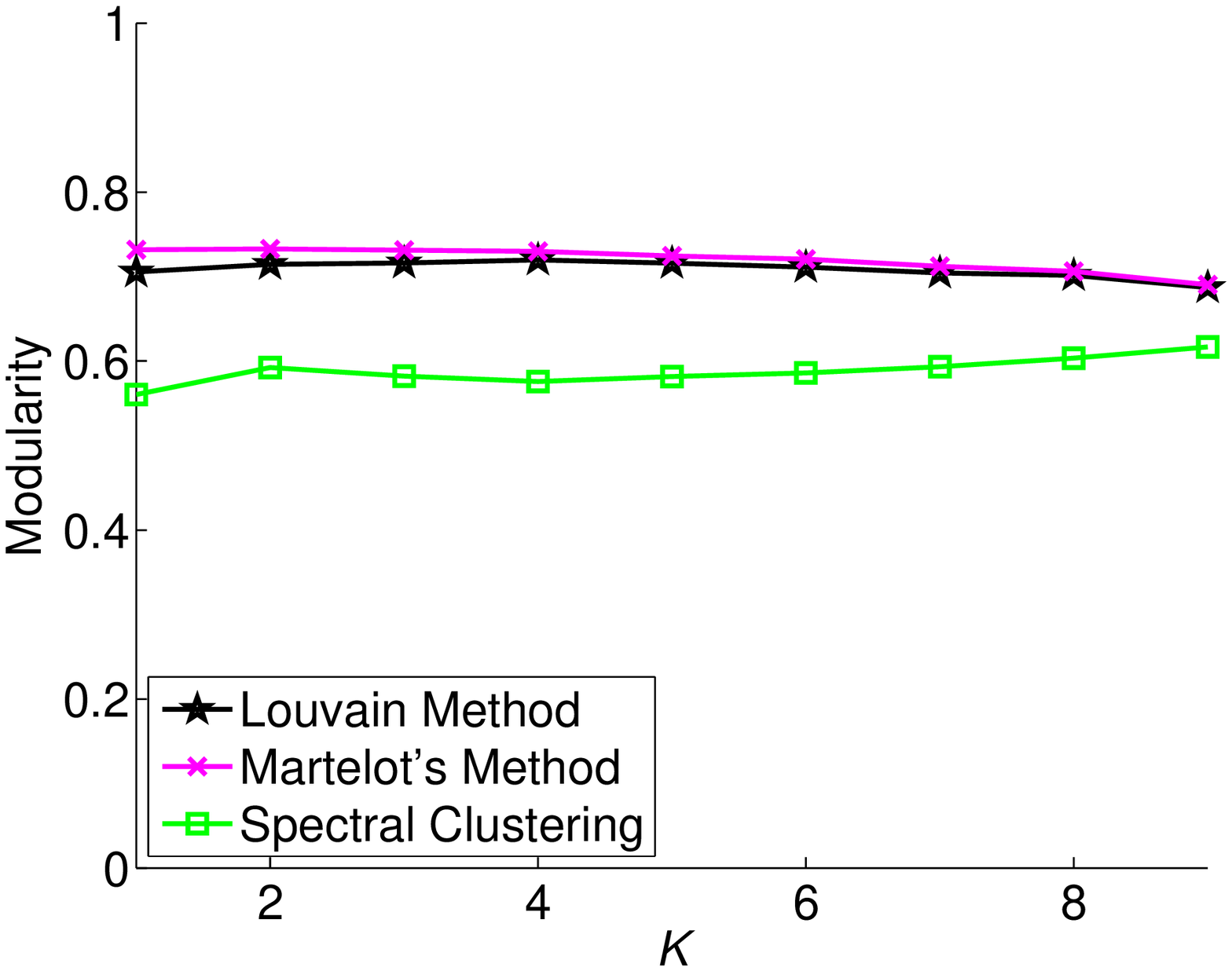}} &
\subfloat[ca-AstroPh]{\includegraphics[width=\figwidthStepAll]{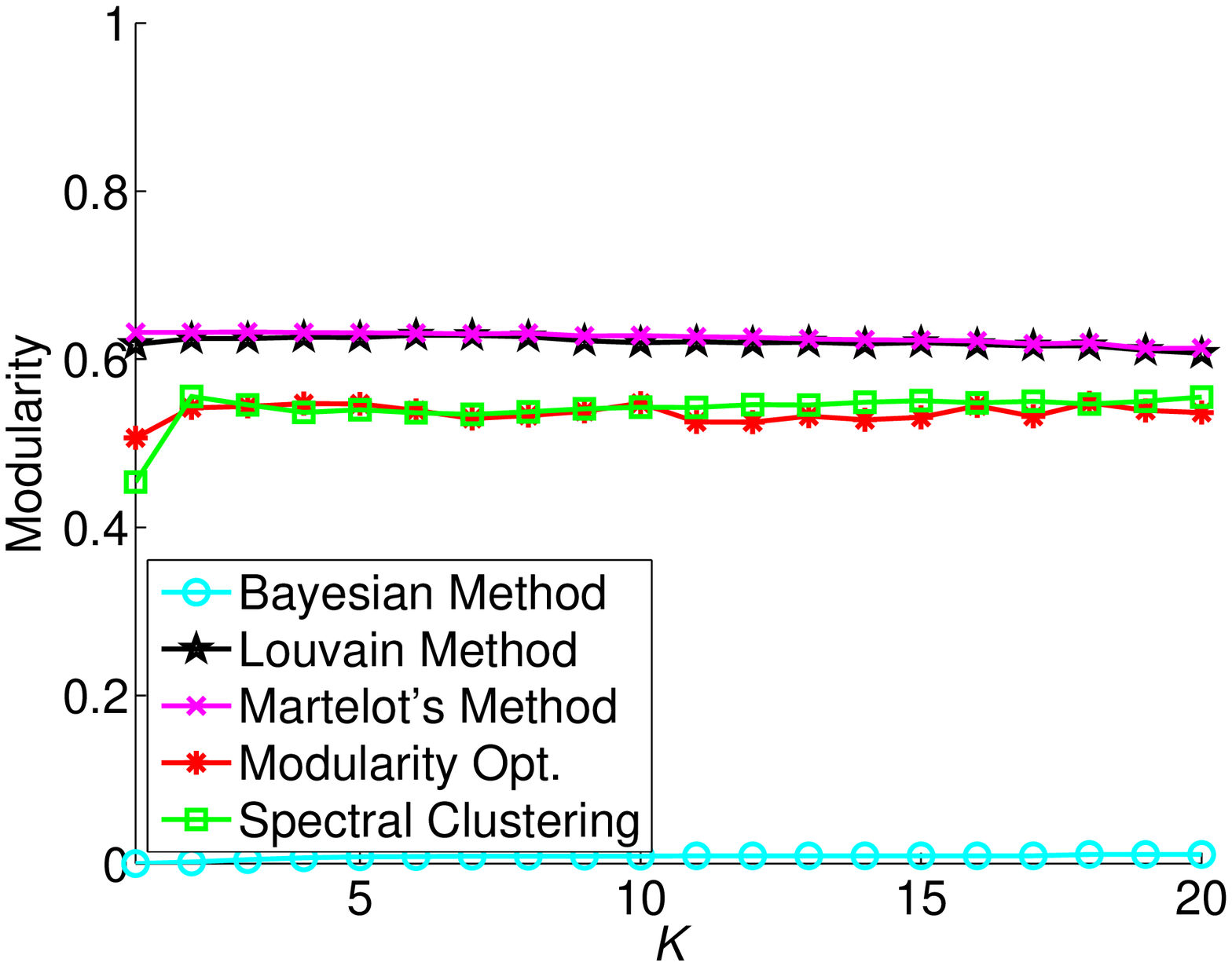}}
&
\subfloat[Email-Enron]{\includegraphics[width=\figwidthStepAll]{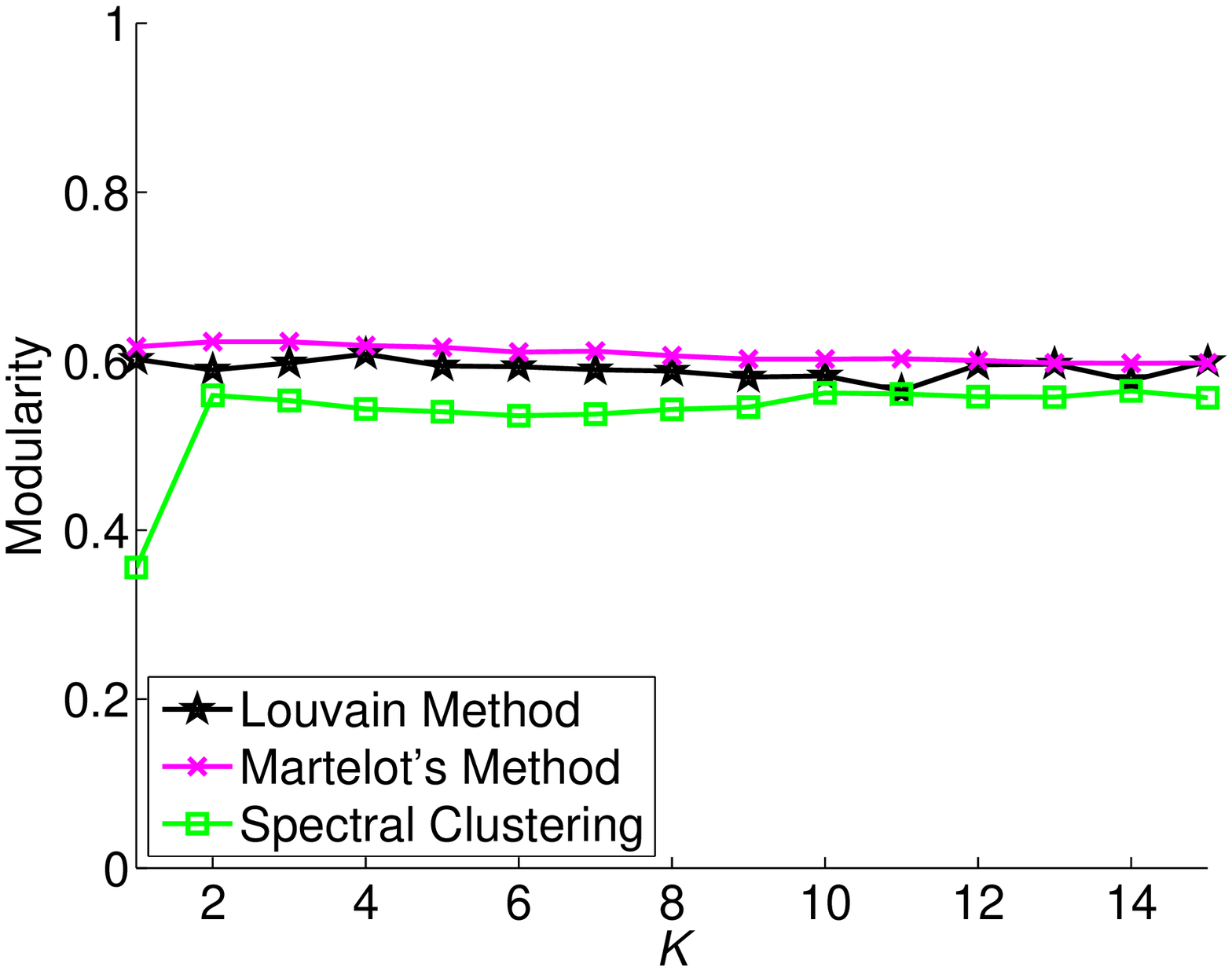}}
 \\
\subfloat[oregon1\_010331]{\includegraphics[width=\figwidthStepAll]{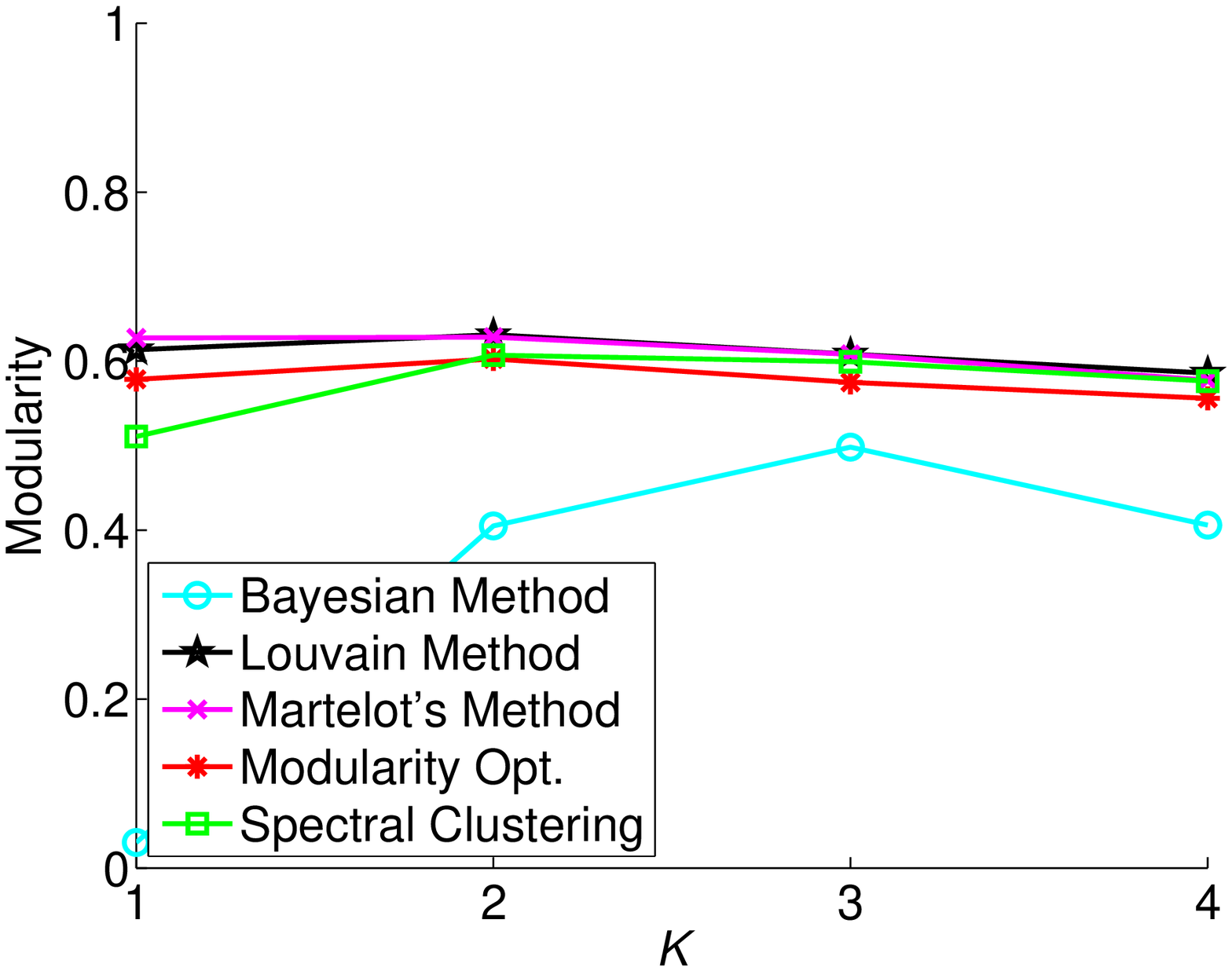}}
&
\subfloat[oregon1\_010421]{\includegraphics[width=\figwidthStepAll]{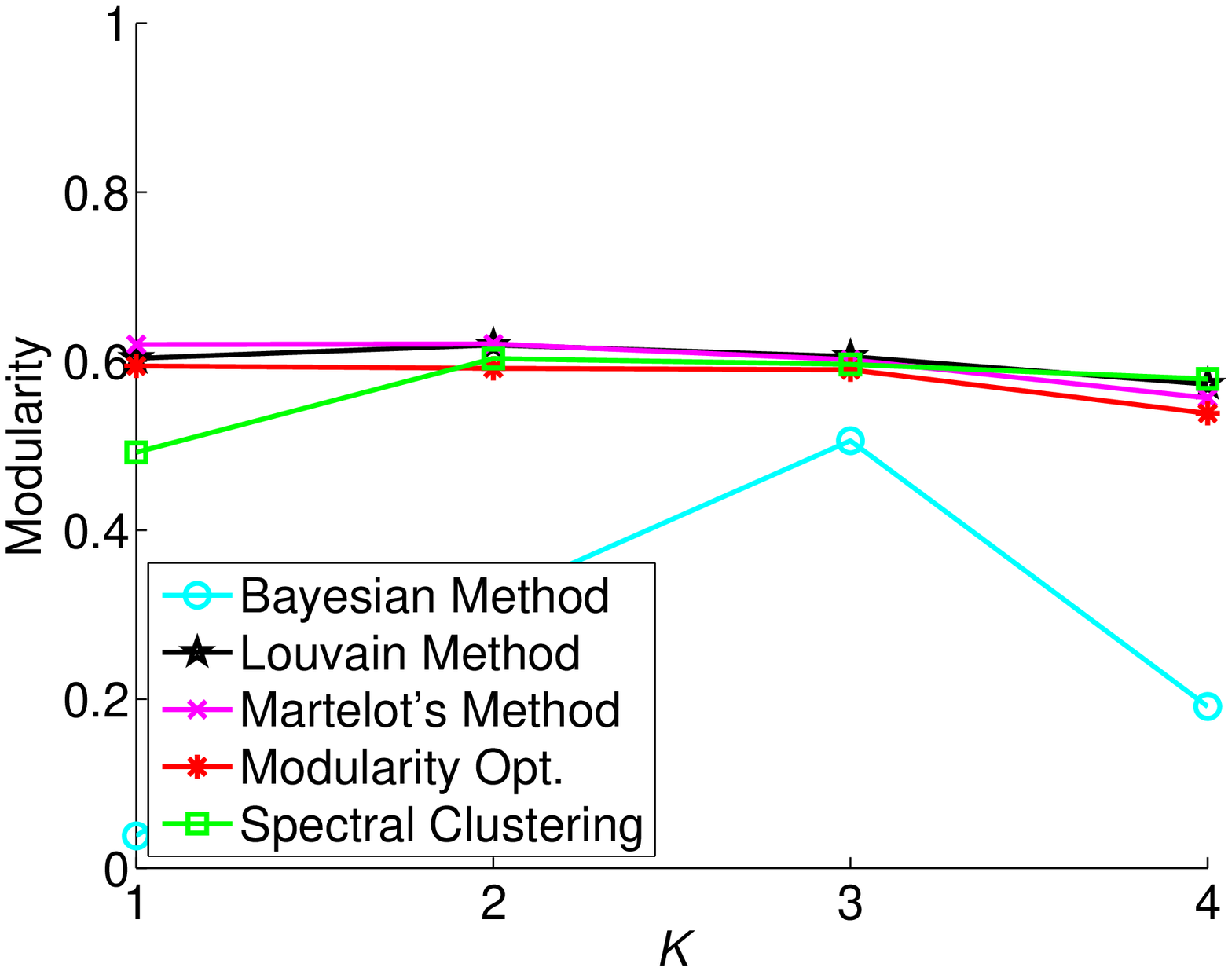}} &
\subfloat[oregon1\_010428]{\includegraphics[width=\figwidthStepAll]{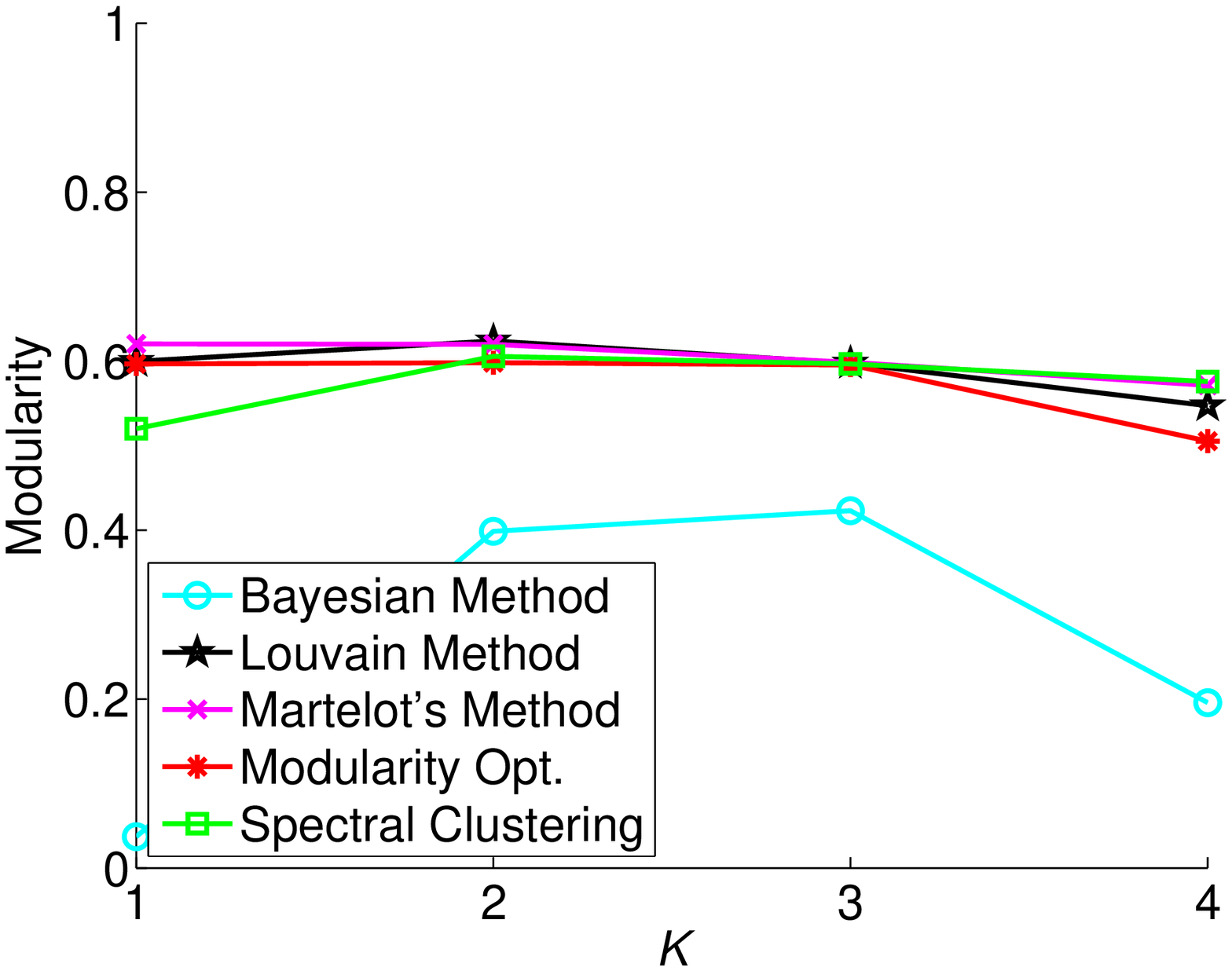}} \\
\end{tabular}
\caption{Modularity over different $K$'s} \label{allModules}
\end{figure*}

\begin{figure*}
\centering
\begin{tabular}{ccc}
\subfloat[ca-GrQc]{\includegraphics[width=\figwidthStepAll]{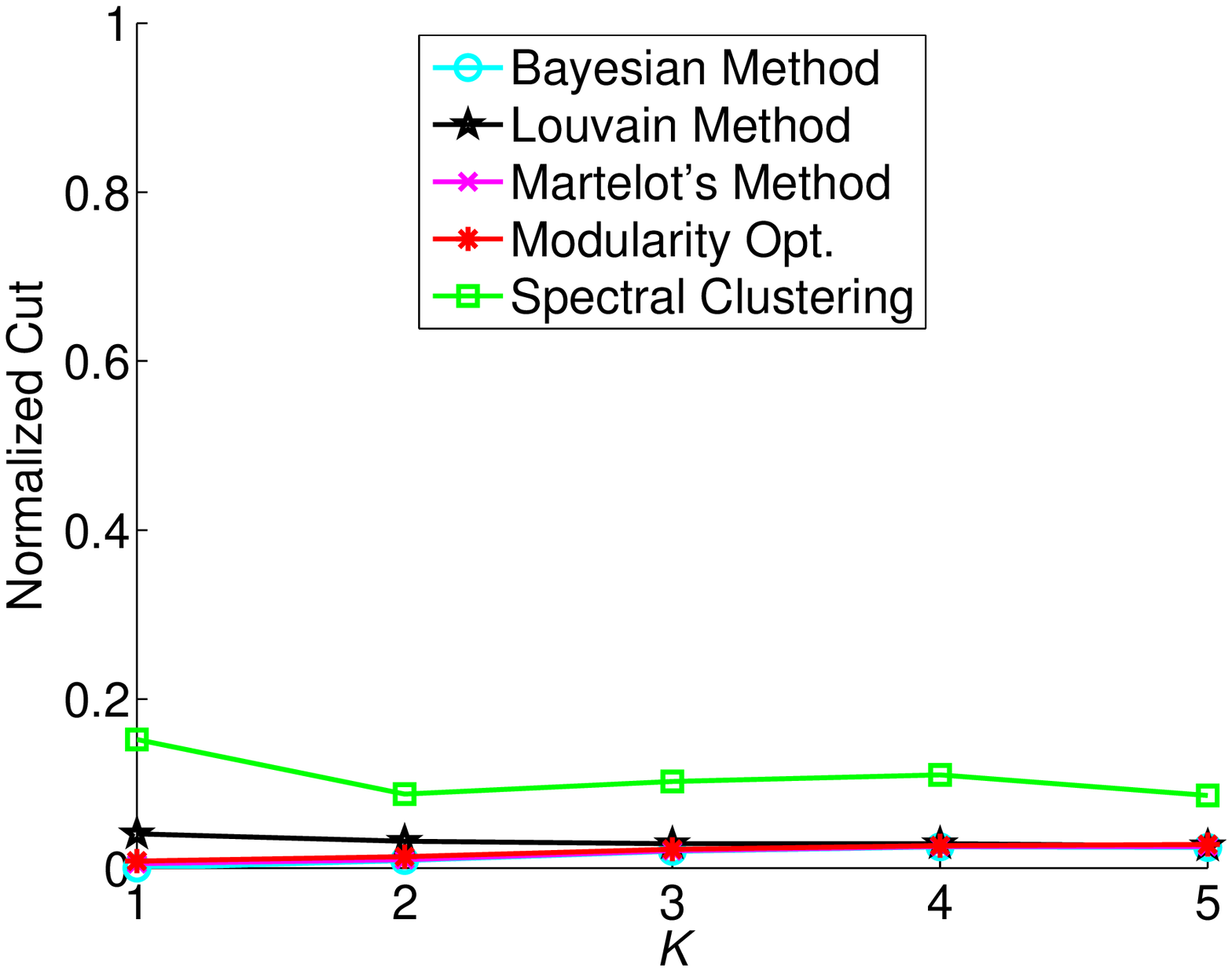}}  &
\subfloat[ca-HepPh]{\includegraphics[width=\figwidthStepAll]{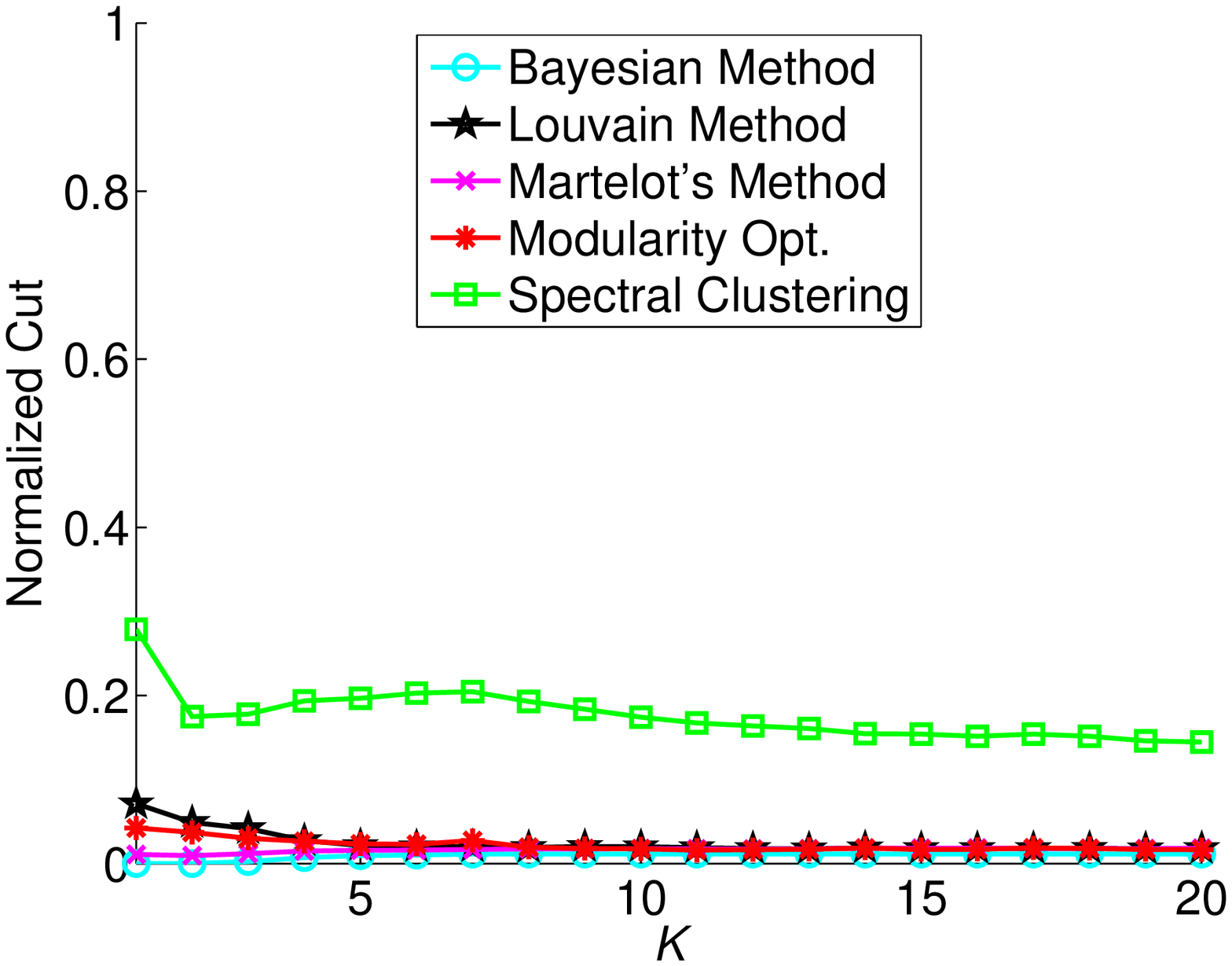}} &
\subfloat[ca-HepTh]{\includegraphics[width=\figwidthStepAll]{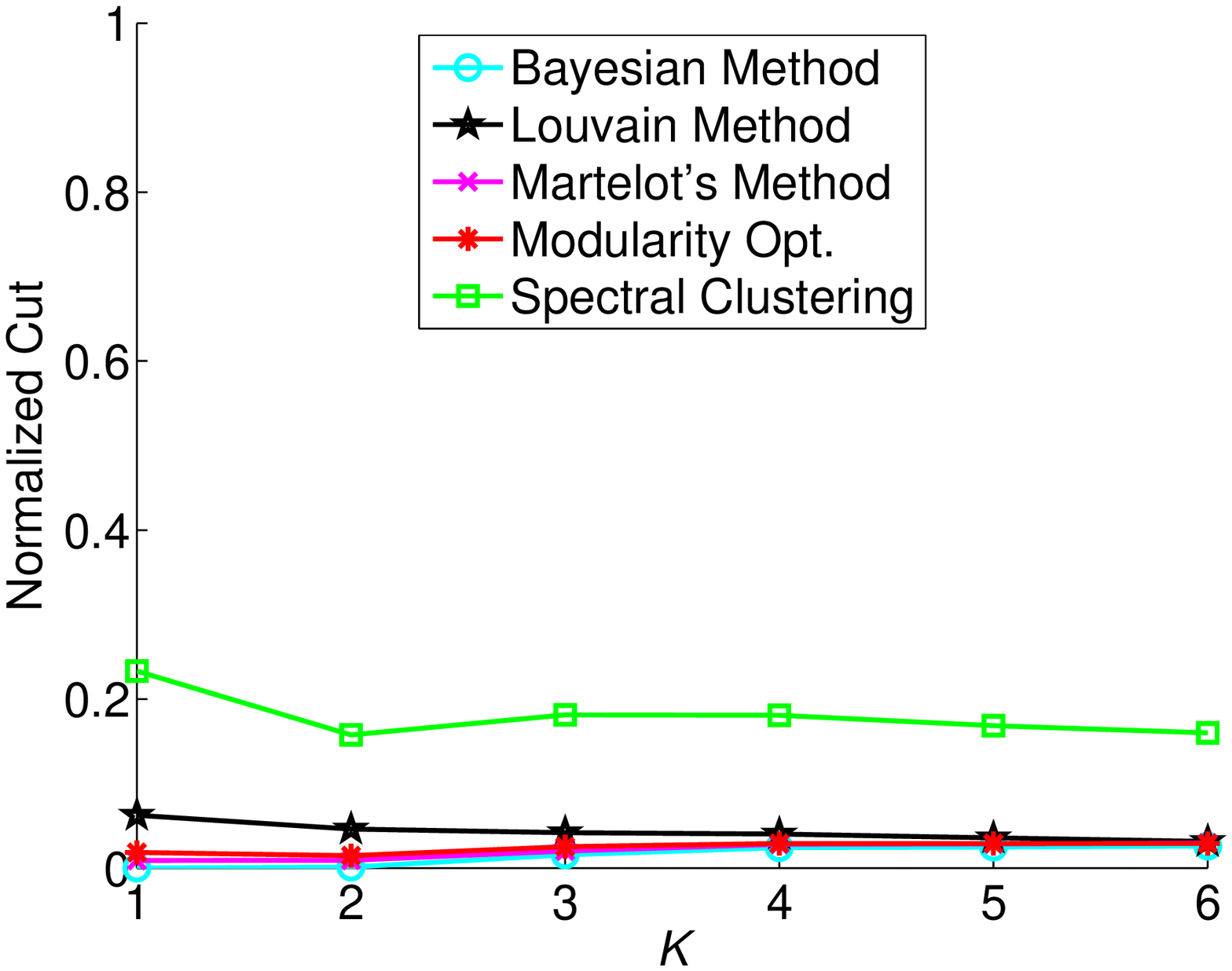}}
\\
\subfloat[ca-CondMat]{\includegraphics[width=\figwidthStepAll]{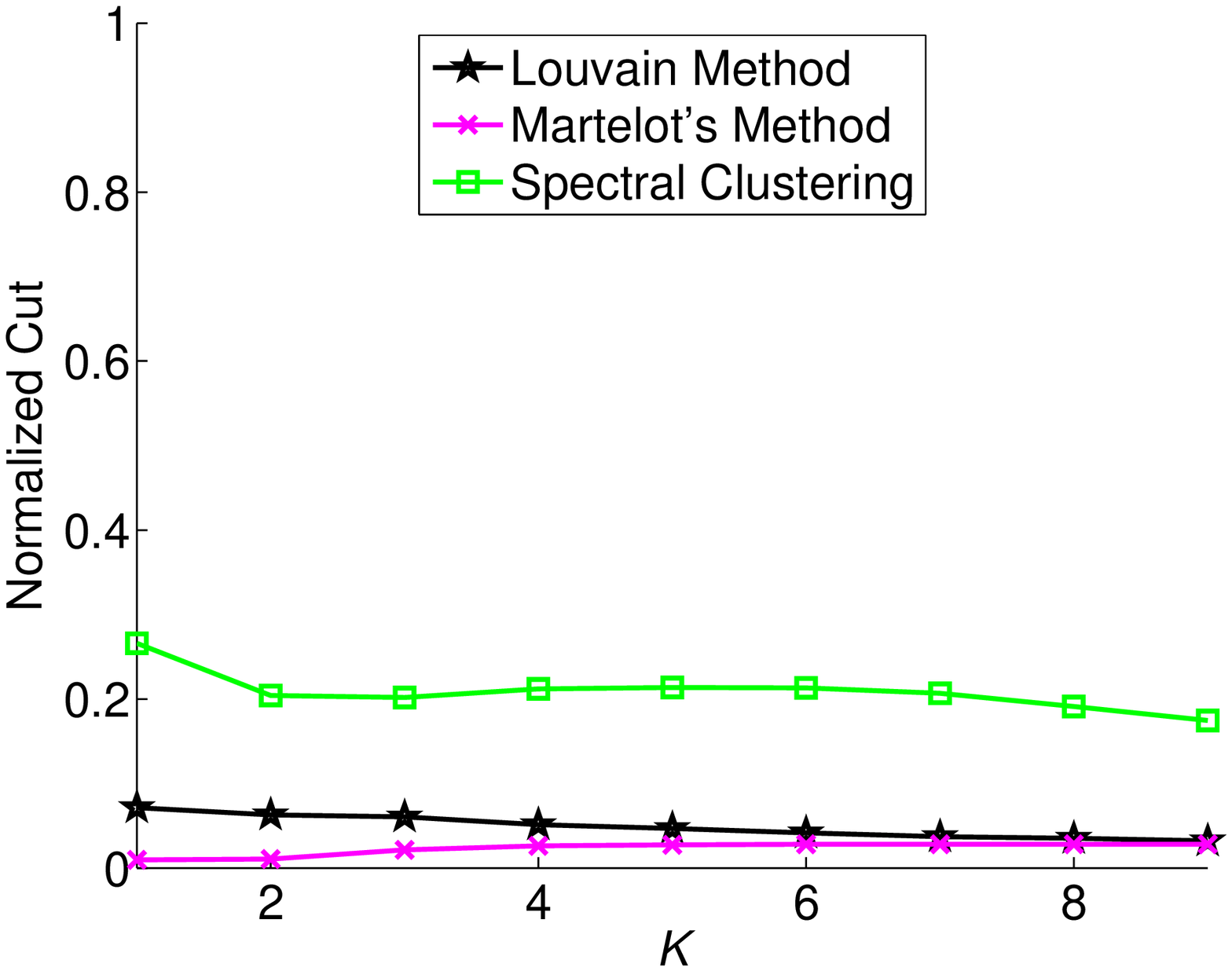}} &
\subfloat[ca-AstroPh]{\includegraphics[width=\figwidthStepAll]{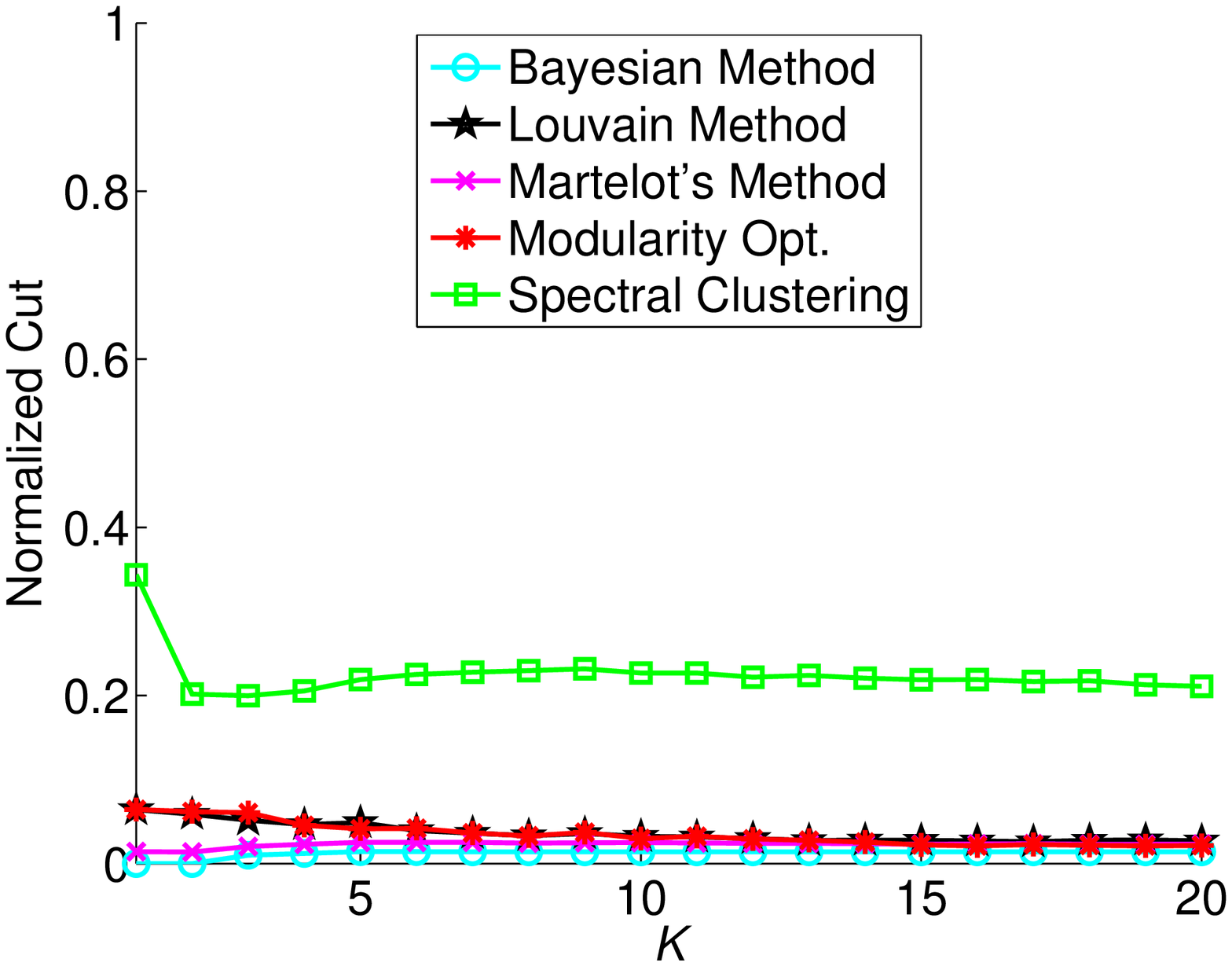}}
&
\subfloat[Email-Enron]{\includegraphics[width=\figwidthStepAll]{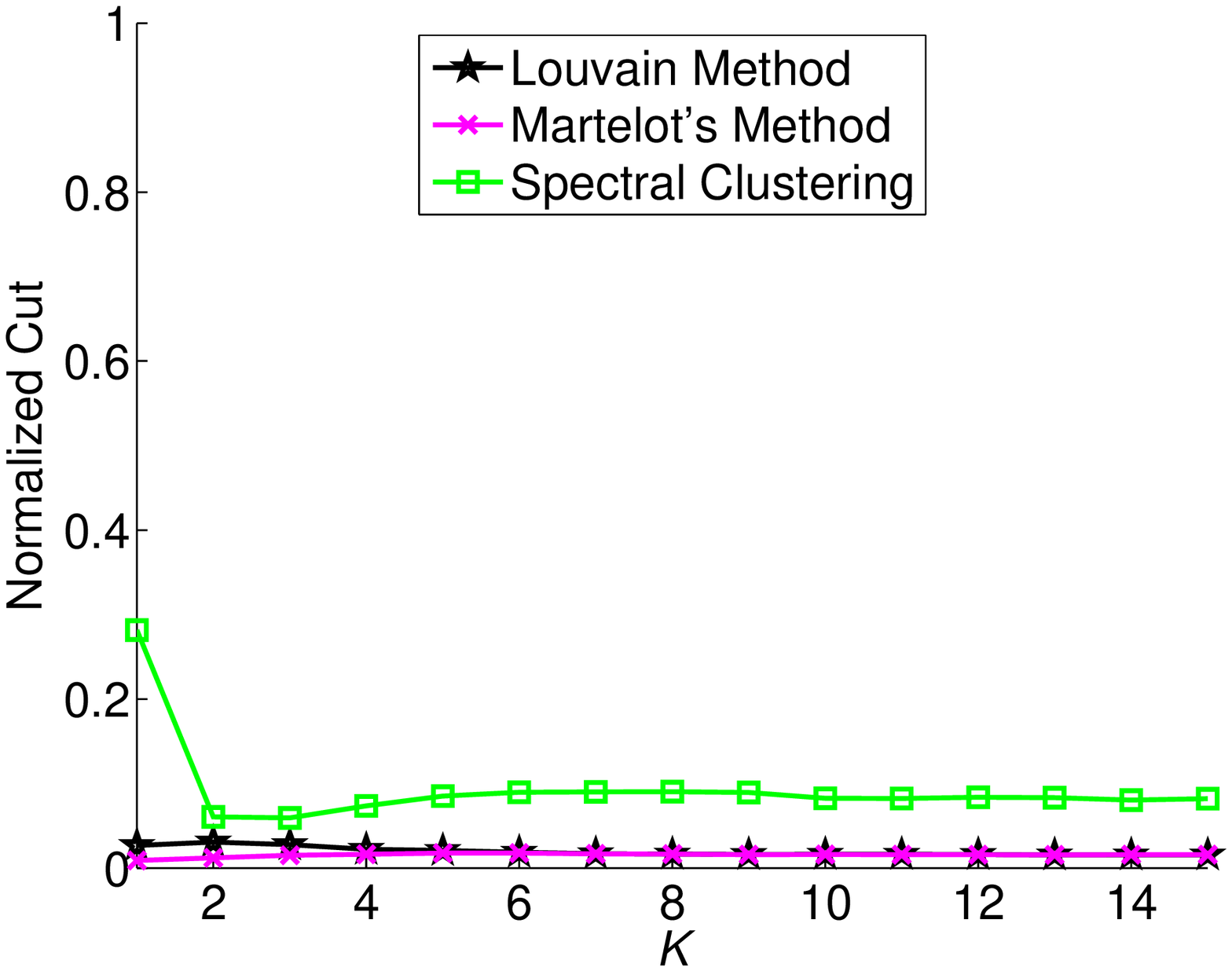}} \\
\subfloat[oregon1\_010331]{\includegraphics[width=\figwidthStepAll]{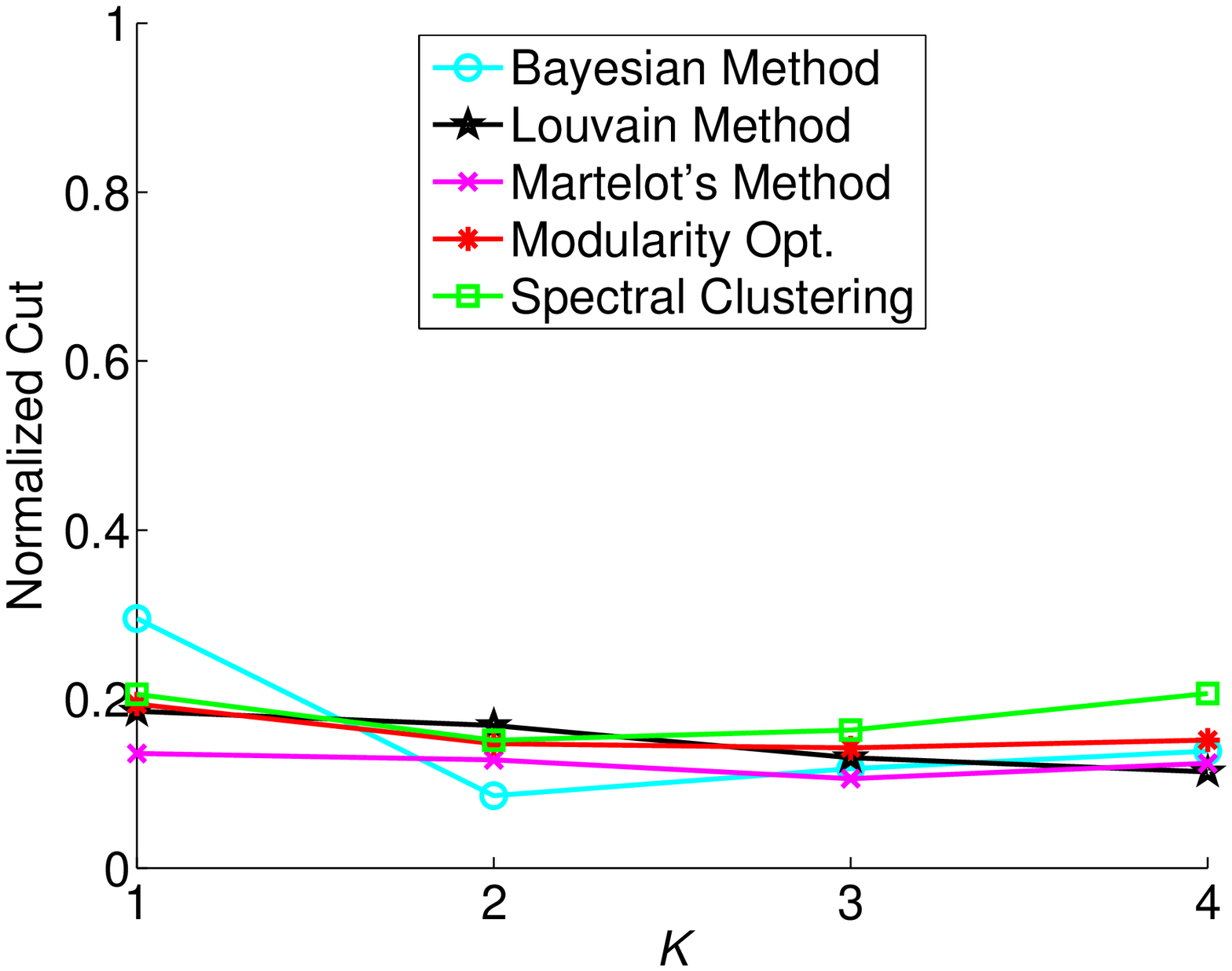}}
&
\subfloat[oregon1\_010421]{\includegraphics[width=\figwidthStepAll]{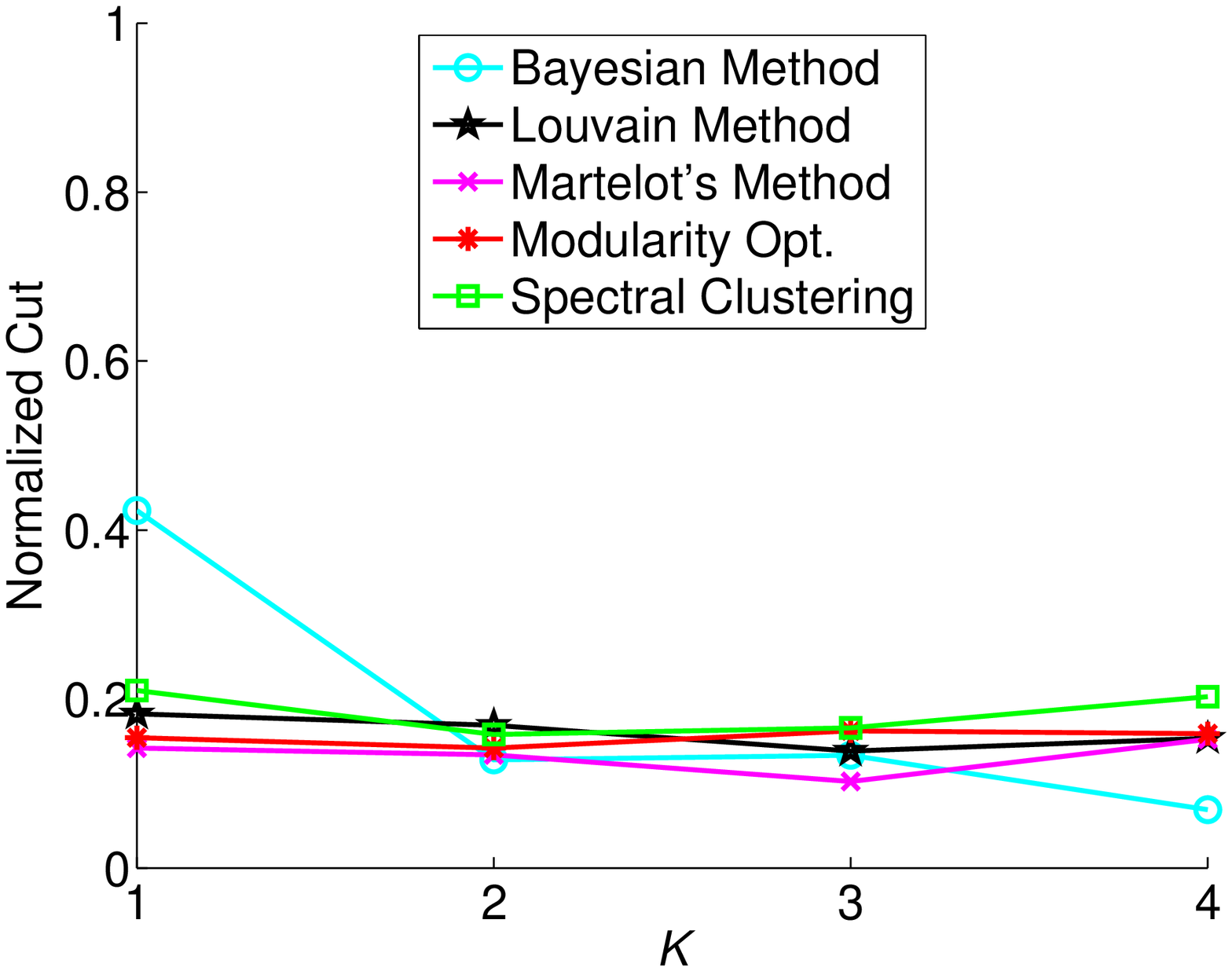}} &
\subfloat[oregon1\_010428]{\includegraphics[width=\figwidthStepAll]{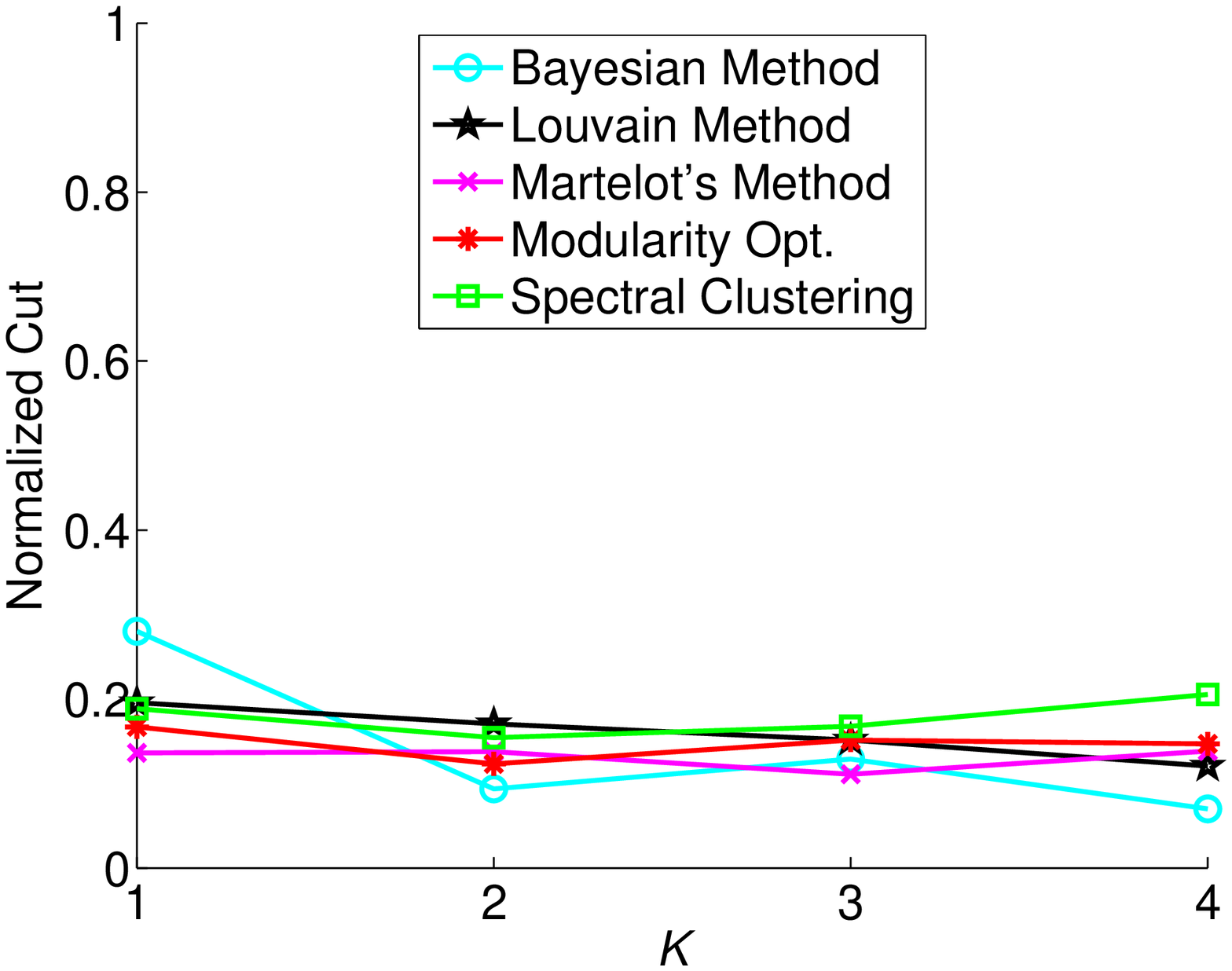}} \\
\end{tabular}
\caption{Normalized cut over different $K$'s} \label{allNormcut}
\end{figure*}

Finally, we consider a collection of results on nine more real-world graphs.
In Fig.~\ref{allModules},  \ref{allNormcut} and \ref{allTime}, label propagation \cite{raghavan2007near}  is omitted  because the quality is usually much worse than other algorithms. Additionally, the Bayesian method \cite{hofman2008bayesian}  and modularity optimization \cite{newman2004fast} are omitted in two comparisons because they run too slowly.

In Fig.~\ref{allTime}, the maximum running time for each algorithm is
normalized to one.
Generally, the overall run time is substantially reduced. The main exception is that the runtime increases for spectral clustering on the oregon networks (which are quite small and sparse).
Otherwise, the running time reduction can be as much as 80\% for $K=2$ and over 95\% for larger values of $K$.
The small fluctuations in running time are due to the same phenomenon as we encountered for Facebook (Fig.~\ref{figFacebook2}) where the number of iterations in the clustering method changes for some reason.

In terms of quality, Fig.~\ref{allModules} shows the modularity as $K$ increases. Generally, the modularity is little changed and sometimes even improves. For instance, the time to run Modularity Optimization on ca-HepTh is over 1000 seconds on the full graph, but that time reduces by 95\% for $K=4$ with no reduction whatsoever in modularity score.

For another quality measure, we show the normalized cut \cite{leskovec2010empirical} in Fig.~\ref{allNormcut} (lower is better). Once again, we see almost no degradation in quality even when runtimes are reduced by 50-90\% or more.

{

\begin {table}[!ht]
\footnotesize
\caption {Remaining Nodes in $K$-core in Multiple Graphs} \label{tableGraphsK}
\begin{center}
\begin{tabular}{l*{3}{c}r}
Data Set Name             & K (Remaining Ratio)\\
\hline
Email-Enron           & 3 (0.58), 6 (0.25), 9 (0.14) &\\
ca-AstroPh &5	(0.66), 10 (0.44), 20 (0.23)\\
ca-CondMat &4 (0.58),  6 (0.35), 8 (0.19)\\
ca-GrQc	& 3	(0.50),   4 (0.30),  5 (0.17)\\
ca-HepPh	& 5 (0.45), 10 (0.23),  20 (0.16) \\
ca-HepTh &3 (0.52), 4 (0.32),  5 (0.21)\\
oregon1\_010331 &2 (0.64), 3 (0.20),  4 (0.08) \\
oregon1\_010421 &2 (0.65),  3 (0.21),  4 (0.09) \\
oregon1\_010428 &2 (0.65),  3 (0.20),  4 (0.08) 
\end{tabular}
\end{center}
\end{table}


Table \ref{tableGraphsK} lists the ratio of remaining nodes in $K$-core. With Fig.~\ref{allModules}, it demonstrates empirically that if the remaining ratio in $K$-core is above 50\%, the modularity obtained by our framework is almost invariant, and if the remaining ratio is above 20\%, the modularity decreases slightly. In all those cases, the running time reduces significantly as the number of remaining nodes decreases.

}

\section{Theoretical Analysis} \label{sectionTheory}

In this section we will discuss the theoretical basis of our proposed method. The essence of our approach is that the dense communities are preserved after a $K$-core reduction, while the  size of the graph reduces drastically.  Subsequently, we can identify the same communities with much less effort. The reduction is  graph size after $K$-core reductions is widely studied and  empirically documented in our experimental results.  But what happens to the communities?

For the analysis, we assume each community contains a subgraph with an Erd\H{o}s-R\'enyi (ER) structure, i.e., all vertices within the ER subgraph are equally likely to be connected. Note that full communities may not have a uniformly random structure because they may contain many nodes of smaller degrees, but for our analysis it is sufficient for a community to just contain a dense ER subgraph of adequate size. Our analysis is based on a beautiful result  by Pittel, Spencer, and Wormald~\cite{PiSp96}. Here we state the key result:
\begin{quote}
For the Erd\H{o}s-R\'enyi random graph $G(n, m)$ on $n$ vertices with $m$ edges, and for $k \geq 3$, with high probability, a giant
$K$-core appears suddenly when $m$  reaches $c_K n/2$; here $c_K = \min_{\lambda >0} \lambda/\pi_K(\lambda)$ and
$\pi_K(\lambda) = P\{{\rm Poisson}(\lambda) \geq K -1\}$.
\end{quote}

\begin{figure}
  \centering
  \includegraphics[width=0.25\textwidth]{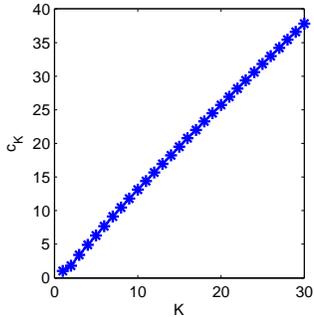}
  \caption{Average degree ($c_K$) that ensures a $K$-core exists in an ER subgraph with high probability.}
  \label{figtheory}
\end{figure}

Note that $c_K$ in this result is equal to the average degree of the ER graph.
This result shows that an ER subgraph will preserve most of its vertices as long as its average degree \emph{within the ER subgraph} is high enough. Figure~\ref{figtheory} shows the  $c_K$ values that ensure that a $K$-core exists with high probability.  As seen in this figure, $c_K$ values are quite realistic, and even for $K=25$ any community with an average degree of around 30 will have a high probability of containing a $K$-core and hence be preserved.

For our full approach, consider a graph as  combination of a random graph (where two edges are connected with a probability directly proportional to the product of the vertex degrees) and a collection of denser ER subgraphs. Note that  the first component, the random graph, is what the modularity metric uses as a null hypothesis, and thus maximizing modularity should be equivalent to identifying the dense ER graphs of the second part.  Consider applying a $K$-core reduction on this graph. By the results of Pittel, Spencer, and Wormald \cite{PiSp96}, we know that  dense communities will be preserved with high probability.  However, we also know that many vertices and edges of the random graph (specifically vertices that are not part of the communities) will be eliminated.  This will result in improved efficiency due to reduced graph size. Moreover, the dense structures will be more pronounced due to eliminations of the noise.

\subsection{Selecting $K$}

This theoretical understanding leads to motivation for choosing $K$. As we can see from our experimental results, tightly knit communities like Facebook and the coauthorship networks can have relatively high values of $K$, like $K=20$. According to our analysis, an ER subgraph with an average degree of 25 (within the community) has a high probability of containing a 20-core, so it is not surprising that some social networks demonstrate this phenomenon. For the coauthorship networks, coauthors of the same paper form a clique, so fields that have papers with high numbers of authors can have higher values of $K$.
On the other hand, values like $K=3$ only require ER subgraphs with average degree (within the subgraph) of 3.35. Hence, if communities are of approximately size 10, then number of edges should be 67 for a high probability of a 3-core.
{ When nodes in each community are degree assortative  \cite{ciglan2013community}, high degree nodes tend to remain in $K$-core, and we can use a larger $K$.

In practice, we can choose  $K$ simply according to the size of the subgraph. For example, we may choose the largest possible $K$ with the $K$-core subgraph containing at least $20\%$ of the total nodes. When nodes in each community are degree assortative  \cite{ciglan2013community}, high degree nodes tend to remain in $K$-core, and we can use a larger $K$.}

\subsection{Node Remaining Ratios in K-core}

\label{subsectionPeelingTheo}

In this subsection, we define the remaining ratio $R$ as the ratio of the nodes in the $K$-core to all the nodes in the graph, and analyze the relationship between $R$ and $K$ .

First, we define several variables: 
$N$, the total number of nodes; 
$K$ is the parameter for the $K$-core subgraph; $\mathcal{V}$ is the set of nodes in a community, and $\mathcal{V}^{(t)}$ is the set of remaining nodes after the $t$th removal by the while loop in Algorithm \ref{algKcore1}; and $y^{(t)}(k)$ is the probability of being degree $k$ for remaining nodes after the $t$th removal. Hence, $t = 0$ indicates the status of the original graph, without any removal. $k_{\max}$ is the maximum possible degree of any node in $\mathcal{V}$. $R^{(t)}$ is the remaining ratio of nodes after $t$th removal, and $R^{(0)}= 1$.

As defined, $y^{(0)}(k)$ should be the degree distribution of the original graph. In each removal, we define the remaining degree $D_{rn}$ introduced by edges between the remaining nodes after this round removal, and the removed degree $D_{rm}$ which is due to the edges between the remaining nodes and the removed nodes by current removal.

In the $t+1$th removal ($t \geq 0$), the total number of remaining degrees is
\begin{equation}
D_{rn}^{(t+1)} = \sum_{k=K}^{k_{\max}} [ k\times y^{(t)}(k)]\times N \times \frac {\sum_{k=K}^{k_{\max}} [ k\times y^{(t)}(k)]}{\sum_{k=0}^{k_{\max}} [ k\times y^{(t)}(k)]}
\end{equation}
where $\sum_{k=K}^{k_{\max}}  y^{(t)}(k) $ represents the expected ratio of neighboring nodes inside $\mathcal{V}^{(t+1)}$. 


The  total number of removed degrees is
\begin{equation}
D_{rm} ^{(t+1)} = \sum_{k=K}^{k_{\max}} [ k\times y^{(t)}(k) ]\times N \times \frac {\sum_{k=0}^{K-1} [ k\times y^{(t)}(k) ] }{\sum_{k=0}^{k_{\max}} [ k\times y^{(t)}(k) ]}
\end{equation}

%
%

Then, for nodes in $\mathcal{V}^{(t+1)}$, the ratio of remaining degree on average after the $t+1$th removal is
\begin{equation}
r^{(t+1)}  = \frac{D_{rn}^{(t+1)}  }{D_{rn}^{(t+1)}   +D_{rm}^{(t+1)} } 
= \frac{\sum_{k=K}^{k_{\max}} [ k\times y^{(t)}(k)]}{\sum_{k=0}^{k_{\max}} [ k\times y^{(t)}(k)]}
\label{EqRemainingDegree}
\end{equation}

For a node of degree $k_c$, if edges are randomly kept at probability $r$, then the probability of having $k$ remaining edges is
\begin{equation}
z^{(t+1)} (k, k_c) = f(k, k_c, r^{(t+1)} ) 
\end{equation}
where $f(k, k_c, r^{(t+1)} )$ is the probability of having $k$ edges successfully kept within a total of $k_c$ edges. Function $f$ is essentially a probability density function of a binomial distribution, and will be used several times in the remaining part of this section.

Therefore, the expected degree distribution after one removal is
\begin{equation}
y^{(t+1)}(k) = \sum_{k_c = K}^{k_{\max}} z^{(t+1)} (k,k_c)y^{(t)}(k_c)
\end{equation}

Finally, we have the expected remaining ratio after $t+1$th removal.
\begin{equation}
R^{(t+1)} =  R^{(t)} \times \sum_{k=K}^{k_{\max}} y^{(t)}(k);
\end{equation}

This procedure can run iteratively as $t$ increases and converge asymptotically as $r$ reduces. It is noticeable that $r^{(t+1)}$  in Eq. (\ref{EqRemainingDegree}) is only determined by the initial degree distribution $y^{(0)}(k)$, so is $R^{(t+1)}$.




\subsection{Maximizing the Modularity in K-core}
 In this subsection, we develop our theorem based on a variation of the planted $l$-partition model \cite{lancichinetti2009benchmarks}. In our model, we assume the nodes in different ground-truth clusters have the same degree distribution and correlation coefficient. For a randomly selected edge, it could be an intra-cluster connection or an inter-cluster connection. If  $\mu$ is defined as the proportion of the first type edges, we assume $\mu$ to be identical for all the ground-truth clusters.

 \begin{proposition}
 Under the assumptions above, if  the modularity  for $G_1$ is maximized by taking the ground-truth clusters represented by $g_1$, then the modularity for $G_K$ is also maximized by taking the corresponding subset $g_K$ from $g_1$.
 \label{prop}
 \end{proposition}

 Let $\rho$  be the remaining ratio of edges in $G_K$. Because each cluster has the same degree distribution and correlation coefficient, when $K$ is given, $\rho$  is invariant in different clusters.

 We define $E_r = 2me_r$ as total degrees introduced by the edges in cluster $r$ and $A_r = 2ma_r$ as the total degrees introduced by all the edges connecting to nodes in cluster $r$.
 Thus, the modularity can be rewritten as
 \begin{align}
 \mod(G_1, g_1) = & \sum_r[\frac{E_r}{2m} - (\frac{A_r}{2m})^2], \nonumber \\
 \mod(G_K, g_K) = & \sum_r[\frac{\rho E_r}{2\rho m} - (\frac{\rho A_r}{2\rho m})^2] = \mod(G, g_0) . \nonumber
 \end{align}

 Thus, using the same configuration, the modularity in $K$-core is the same as in 1-core. As $g_1$ maximizes the modularity, any other configurations $g'$  can be considered as a series of relabeling certain nodes from one cluster to the another based on $g_1$, which will reduce the modularity. Without loss of generality, we demonstrate that relabeling $n_{\delta}$ nodes degrees from their ground-truth cluster `Cluster 1' to a new cluster `Cluster 2' will reduce the modularity. $A_{\delta}$ is the corresponding number of degrees in $K$-core associated with those nodes.

 After relabeling, the change of intra-cluster degrees is
 \begin{align}
 & -\frac{\mu A_{\delta}\frac{\rho A_1 - A_{\delta}}{\rho A_1}}{2\rho m} + \frac{ (1 - \mu)A_{\delta}\frac{ \rho A_2}{\sum_r \rho A_r}}{2\rho m},  
 \label{modchange_eq1}
 \end{align}
 where $\mu A_{\delta}$ is the intra-cluster degree introduced by edges connecting to the  relabeling nodes in ground-truth Cluster 1, and  the first term means the loss is proportional to the remaining degrees in Cluster 1. The second term means the degree increase is proportional to the degrees in Cluster 2.
 On the other hand, the change caused by the null model is
\begin{eqnarray} 
 & -[(\frac{\rho A_1 - A_{\delta}}{2\rho m})^2 + (\frac{\rho A_2 + A_{\delta}}{2\rho m})^2 ] \nonumber  \\
& + [(\frac{\rho A_1 }{2\rho m})^2 + (\frac{\rho A_2 }{2\rho m})^2 ]  \nonumber \\
 = & \frac{A_{\delta}}{2\rho m} [ \frac{\rho A_1 - \rho A_2 -  A_{\delta}}{\rho m} ].
 \label{modchange_eq2}
\end{eqnarray} 

 Therefore, the total change is the summation of Eq. (\ref{modchange_eq1}) and (\ref{modchange_eq2})
 \begin{align}
 &\mod(G_K, g_K') -  \mod(G_K, g_K)   \nonumber \\
 = &  \frac{A_{\delta}}{2\rho m}[\mu \frac{\rho A_1 - A_{\delta}}{\rho A_1} + (1 - \mu)\frac{  A_2}{\sum_r  A_r} \nonumber \\
 &   +  \frac{\rho A_1 - \rho A_2 -  A_{\delta}}{\rho m}  ] \nonumber \\
 = &  \frac{\frac{A_{\delta}}{\rho}}{2\rho m}[\mu \frac{ A_1 - \frac{A_{\delta}}{\rho}}{ A_1} + (1 - \mu)\frac{  A_2}{\sum_r  A_r}  \nonumber  \\
 & +  \frac{ A_1 -  A_2 - \frac{ A_{\delta}}{\rho}}{ m}  ] \nonumber \\
 = &\mod(G_1, g_1') -  \mod(G_1, g_1)   \nonumber \\
 \leq & 0
 \end{align}
 where $g_1'$ corresponding to the configuration of changing  labels of node with $ \frac{ A_{\delta}}{\rho}$ edges from Cluster 1 to Cluster 2. $\rho = 1$ corresponds to the graph $G_1$.

 Thus, the proposition is proved. Therefore, if the Step 2 algorithm can maximize the modularity, the computed community for $K$-core is the same as the ground-truth.  As our Step 3 algorithm also maximizes the modularity for the remaining nodes, the ground-truth community structure of the whole graph can be found by our framework.


\section{Conclusions}\label{sectionConclusions}

In this work, we introduce a $K$-core based framework that can accelerate community detection algorithms significantly. It includes three steps: first, find the $K$-core of the original graph; second, run a community detection algorithm on the $K$-core; third, find labels for all the nodes outside the $K$-core and then optimize for the whole graph. Experiments demonstrate efficiency and accuracy on different real graphs, under several quality measurements. Theoretical analysis supports our approach through $K$-core. 
The MATLAB code will be made freely available for download. For future work, we consider the case of overlapping communities.

\bibliographystyle{siam}
\bibliography{kcore3.bib}


\end{document}